\documentclass[apl, twocolumn]{revtex4-1}
\usepackage{graphicx}
\usepackage{dcolumn}
\usepackage{bm}
\usepackage{epstopdf}
\usepackage{color}
\usepackage{ulem}
\begin{document}

\title
{Defect complexes in Ti-doped sapphire: A first  principles study.}
\author{L.\,Yu.\,Kravchenko$^1$, D.\,V.\,Fil$^{1,2}$}
\email{fil@isc.kharkov.ua}
\affiliation{$^1$Institute for Single Crystals, National Academy of
Sciences of Ukraine, 60 Nauky Avenue, Kharkiv 61001, Ukraine\\
$^2$ V.N. Karazin Kharkiv National University, 4 Svobody Square, Kharkiv 61022, Ukraine}

\begin{abstract}
First-principles calculations have been performed to study the formation of defect
complexes in Ti doped $\alpha$-Al$_2$O$_3$ crystals. The formation energies of isolated
Ti$^{3+}$ and Ti$^{4+}$ defects, pairs, triples and quadruples of Ti ions and Al
vacancies are computed under different equilibrium conditions of Al-Ti-O related phases.
Taking into account charge neutrality of the whole system we determine the equilibrium
concentrations of simple and complex defects as well as the total equilibrium
concentration of Ti in an $\alpha$-Al$_2$O$_3$ crystal. It is shown that the equilibrium
concentration of complex defects can be on the same order of or even larger than the
concentrations of isolated substitutional Ti$^{3+}$ and Ti$^{4+}$ defects. It is found
that in Ti-deficient conditions the relative fraction of isolated defects increases and the
balance is shifted towards Ti$^{4+}$ defects. A universal relation between equilibrium
concentrations of isolated and complex defects is obtained. The band structure of the system
with complex defects is calculated and extra
 levels inside the band gap
caused by such defects are found.
\end{abstract}


\maketitle

\section{Introduction}

Doping of synthetic crystals  with different activating ions  provides
      desired optical and lasing properties of such materials \cite{1}. Activating ions can be in
      different charge
      states, may occupy different crystallographic positions and  form complexes of two or
      more impurity atoms situated close to each other or complexes of impurity atoms
      with intrinsic defects.

      In particular, such a situation is realized in Ti:sapphire in which Ti ions  can exist
      in different charge states  and may form pairs, triples  and multisite clusters.
       Ti:$\alpha$-Al$_2$O$_3$  is known as  a laser material \cite{2,3,4,5,6}.
The laser efficiency of Ti:sapphire is affected by a residual infrared absorption in the
emission band of the Ti:Al$_2$O$_3$ laser. Crystal field calculations \cite{7} support a
hypothesis \cite{5} that the absorption is caused by $\mathrm{Ti}^{3+}-\mathrm{Ti}^{4+}$
pairs. The ratio of the absorption coefficients at the pump wavelength ($\lambda=514$ nm)
and at the maximum of residual absorption ($\lambda=800$ nm) is known as the figure of
merit (FoM) of Ti:sapphire laser crystals. FoM can be used to evaluate experimentally the
 concentration of Ti$^{4+}$ ions \cite{8}. The Ti$^{4+}$ concentration can also be obtained from
ultraviolet absorption spectra \cite{9}.

 Defect formation energies and
relative stability of defects in different charge states can be determined from the
first-principles calculations. The formation energy depends on the oxygen chemical
potential $\mu_\mathrm{O}$ and the electron Fermi energy $E_F$. In \cite{10} the
formation energies of intrinsic (native) defects in pure
$\alpha-\mathrm{Al}_2\mathrm{O}_3$ were obtained using the plane-wave pseudopotential
method. Considering various charge states of the defects, the authors of \cite{10}
found that the defect species in their highest charge state exhibit the smallest
formation energies. According to \cite{10}, in a wide range of
 $\mu_\mathrm{O}$ the formation energies of charged vacancies ($V$) and interstitial ions ($i$)
are in the order of $V_\mathrm{Al}^{3-}<O_i^{2-}<V_\mathrm{O}^{2+}<\mathrm{Al}_i^{3+}$.
The formation energies of Schottky and Frenkel defects were also calculated in \cite{10}.
These energies are independent of $\mu_\mathrm{O}$. It was found that the Schottky
quintet has the formation energy per defect smaller than O Frenkel and Al Frenkel pairs
have. In addition, the formation energy of Al Frenkel pairs is smaller than that of O
Frenkel pairs.

The dependence of the formation energies of native defects in $\alpha$-Al$_2$O$_3$  on
$\mu_\mathrm{O}$ and $E_F$  was found in \cite{11}. It was shown  that over most of the
range of $E_F$ the defects in their highest charge states dominate. Nevertheless a
significant amount of defects in lower charge states can emerge at some $\mu_\mathrm{O}$
and $E_F$. In particular, the formation energy of the oxygen vacancy $V_\mathrm{O}^{1+}$
can be comparable to the energy of other defects. This result correlates with the
experimental observation of F+ centers in $\alpha$-Al$_2$O$_3$ \cite{12}. Calculations
\cite{11} support the conclusion of \cite{10} that for the charge neutral combinations  the
formation energy ordering is Schottky $<$ Al Frenkel $<$ O Frenkel.

The charge states and formation energies of vacancies, interstitial and antisite atoms in
pure $\alpha$-Al$_2$O$_3$ were studied in \cite{13}. It was found that under O-rich and
O-deficient conditions the most stable defect is the Al vacancy $V_\mathrm{Al}^{3-}$.
According to \cite{13} the preferable charge state of O vacancies is $V_\mathrm{O}^0$. In
O-rich conditions the formation energies are ordered as $V_\mathrm{Al}^{3-} <
\mathrm{O}_\mathrm{Al}^{3-} < \mathrm{O}_i^{2-} < V_\mathrm{O}^0 < \mathrm{Al}_i^{3+} <
\mathrm{Al}_\mathrm{O}^{3+}$, and in O-deficient conditions, as $V_\mathrm{Al}^{3-} <
\mathrm{O}_i^{2-} < \mathrm{Al}_i^{3+} <\mathrm{O}_\mathrm{Al}^{3-} < V_\mathrm{O}^0 <
\mathrm{Al}_\mathrm{O}^{3+}$, where $\mathrm{O}_\mathrm{Al}$ and $\mathrm{Al}_\mathrm{O}$
correspond to  O and Al antisite atoms, respectively. The energy ordering for neutral
combinations of native defects is Schottky defect $<$ cation Frenkel $<$ anion Frenkel
$<$ antisite pair. The charge states of the components of the most stable Schottky and
Frenkel defects in \cite{13} differ from one obtained in \cite{10,11}.

In \cite{14}  the energetics of point defects in Ti-doped $\mathrm{Al}_2\mathrm{O}_3$ was
studied. Substitutional and interstitial Ti ions with charge compensating intrinsic
defects were considered and their formation energies against the oxygen chemical
potential were calculated. It was found that substitutional Ti$^{4+}$ ions with charge
compensating Al vacancies are the most stable defects in the oxidized conditions. In
contrast, the formation energy of substitutional Ti$^{3+}$ ions is minimal in the reduced
conditions. In the intermediate range of the oxygen potential the substitutional
Ti$^{3+}$ and Ti$^{4+}$ ions exhibit similar formation energies.

In \cite{15} the formation of Ti clusters in Ti-doped $\mathrm{Al}_2\mathrm{O}_3$ was
investigated. It was shown that Ti$^{3+}$ clusters  have a positive binding energy. The
binding energy  increases with decreasing the distance between Ti$^{3+}$ ions and with
increasing the number of ions in the cluster. It was also found that the binding energy
of a complex of a substitutional $\mathrm{Ti}^{4+}$ ion and an Al vacancy
$V_{\mathrm{Al}}^{3-}$ is rather large.

In this paper we  calculate equilibrium concentrations of  isolated and complex defects
in Ti-doped $\alpha$-Al$_2$O$_3$. A positive binding energy does not automatically mean
that all isolated defects bind
 in clusters. The equilibrium concentrations of  defects  correspond to the minimum
 of free energy. The free energy contains the entropy term and the clustering
 results  in lowering of  entropy.  In the general case both the complex
 and isolated defects are present.
 We find that the equilibrium concentration of  a given complex defect specie
  is proportional to the product of
 the equilibrium concentrations of simple defects which form the complex defect.
  The coefficient of proportionality is determined by the binding
 energy and the temperature.
 The first principles calculations show that the concentration of
  $\mathrm{Ti}^{3+}-\mathrm{Ti}^{3+}$ pairs can be on the same order of or even larger than the
   concentration of isolated $\mathrm{Ti}^{3+}$, while the relative amount of
     $\mathrm{Ti}^{3+}-\mathrm{Ti}^{3+}-\mathrm{Ti}^{3+}$ triples is  small over
     the entire range of allowed $\mu_\mathrm{O}$.

 The  minimum of free energy of the system with charged defects should be found under
 the  additional condition that the overall charge of the defects is equal to zero. It is
 dictated by charge neutrality of the system.  We do not  consider any particular charge
  compensating defects. Charge
neutrality requires that all negatively charged defects compensate all positively charged
defects or vice versa. Applying this condition we find that the equilibrium
concentrations of charged defects are determined by the formation energies of
electrically neutral combinations of charged defects. These combinations can be chosen in
an arbitrary way. The formation energies for electrically neutral combinations of defects
do not depend on the Fermi energy. Therefore, the equilibrium concentrations of charged
defects are independent of $E_F$ irrespectively to the dependence of their formation
energies on $E_F$. We obtain that in addition to $\mathrm{Ti}^{4+}$ isolated defects a
great amount of $\mathrm{Ti}^{4+}-V_\mathrm{Al}^{3-}$ pairs,
$\mathrm{Ti}^{4+}-V_\mathrm{Al}^{3-}-\mathrm{Ti}^{4+}$  triples and
$\mathrm{Ti}^{4+}-\mathrm{Ti}^{4+}-\mathrm{Ti}^{4+}-V_\mathrm{Al}^{3-}$ quadruples
emerge.
 The
maximum concentration of $\mathrm{Ti}^{3+}-\mathrm{Ti}^{4+}$ pairs  is reached at
intermediate values of the oxygen chemical potential and over the entire range of $\mu_O$
this concentration is smaller than the concentration of isolated $\mathrm{Ti}^{3+}$
defects.

It is shown that the ratio of the concentrations of different defect species is changed
in Ti-deficient conditions. At such conditions  most of Ti ions enter into the crystal in
the form of isolated $\mathrm{Ti}^{3+}$ and $\mathrm{Ti}^{4+}$ substitutional defects,
and the balance between $\mathrm{Ti}^{3+}$ and $\mathrm{Ti}^{4+}$ is shifted towards
$\mathrm{Ti}^{4+}$.

The band structure of the system with isolated and complex Ti defects is calculated.
 It is found that the clustering of defects
 results in a shift of defects levels with respect to the
 valence band maximum and in splitting of the
  levels.

\section{Method and computational details}

     Calculations  were performed by the pseudopotential method with the use of a strictly
      localized
atom-centered basis set as implemented in the open source SIESTA code  \cite{16} based on
the density-functional theory (DFT) approach. The pseudopotentials were generated with
the improved Troullier-Martins scheme. The O -- $2s^2 2p^4$, Al -- $3s^2 3p^1$ and Ti --
$4s^1 3d^3$ electron states were considered as  valence configurations and a small core
correction was applied. The generalized gradients approximation with the
Perdew-Burke-Ernzerhof (PBE) exchange-correlation functional  and the double-zeta basis
set plus polarization orbitals was employed.  Primitive translation vectors were allowed
to relax until the maximum residual stress component converged to less than $0.1$ GPa and
atomic positions were optimized until the residual forces had been less than $0.01$
eV/$\mathring{A}$. A real-space grid with the plane-wave cutoff energy $E_c=250$ Ry was
used to calculate the total energy of the system. Selective tests showed that the total
energy was converged within 0.68\ meV/atom for the total energies obtained at $E_c=300$
Ry. In view of large size of the supercell we performed numerical integration over the
Brillouin zone only at the $\Gamma$ point.

     To calculate the formation energy of a given defect a supercell of four fully
     optimized unit
cells ($4\times 30$ atoms) was built. One isolated  or complex defect was placed in the
supercell and the optimization of atomic positions was fulfilled again. The total energy
was calculated for the optimized defect supercell.

     The formation energy of a defect $E_i$ is determined by the difference \cite{17}
     \begin{equation}\label{1}
     E_i = E_{def} - E_{perf} + n_\mathrm{Al} \mu_{\mathrm{Al}}
      +n_\mathrm{O} \mu_\mathrm{O} - n_{\mathrm{Ti}}
     \mu_{\mathrm{Ti}} + q_i E_F,
     \end{equation}
     where $E_{def}$ and $E_{perf}$ are the total energies of
the defect and perfect supercell, correspondingly,
     $n_\mathrm{Al}$ and $n_\mathrm{O}$ are the numbers of Al and O atoms
      removed from the perfect supercell,
$n_{\mathrm{Ti}}$ is the number of Ti atoms added, $\mu_{\mathrm{Al}}$,
$\mu_{\mathrm{O}}$ and $\mu_{\mathrm{Ti}}$ are the chemical potentials of Al, O and Ti,
correspondingly, and $q_i$ is the defect charge in elementary charge units.

\begin{figure}[h]
     \center{\includegraphics[width=0.4\textwidth]{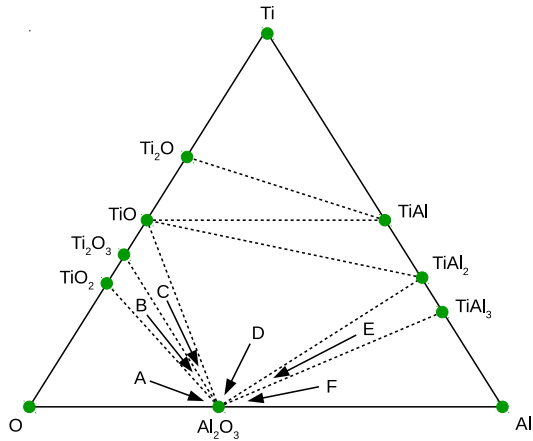}}
     \caption{Computed phase diagram of the Al-Ti-O ternary system. The points A - F
     denoted by arrows correspond to the vertices of
the three-phase region around Al$_2$O$_3$.}\label{f1}
     \end{figure}

     The chemical potentials of Al and Ti vary depending on equilibrium
     conditions of a multiphase  $\mathrm{Al}-\mathrm{Ti}-\mathrm{O}$
ternary  system. To determine these conditions we calculated the formation energies of
$\alpha$-Al$_2$O$_3$, TiO$_2$, Ti$_2$O$_3$,
 TiO, Ti$_2$O, TiAl, TiAl$_2$ and TiAl$_3$ crystals
and  built  the $\mathrm{Al}-\mathrm{Ti}-\mathrm{O}$ phase diagram.
 Calculations were done by the same method with
the same pseudopotentials as ones used for the obtaining of the supercell energies
$E_{def}$ and $E_{perf}$.  The formation energy for a crystal with the general formula
Al$_x$Ti$_y$O$_z$ is given by the equation
      \begin{equation}\label{5}
      \Delta H_f  = \mu_{\mathrm{Al}_x\mathrm{Ti}_y\mathrm{O}_z }
      - x \mu_{\mathrm{Al}} -y \mu_{\mathrm{Ti}} -
      \frac{z}{2}\mu_{\mathrm{O}_2},
      \end{equation}
      where $\mu_\mathrm{Al}$ and $\mu_\mathrm{Ti}$ are the energies per atom
      in the metallic phases, $\mu_{\mathrm{O}_2}$ is the internal energy of
      O$_2$   molecule.
Lattice parameters  obtained by
               structure optimization calculations are presented
         in  Table \ref{tab1}. For comparison, experimental
                lattice parameters are also given in Table \ref{tab1}.
                The computed formation energies are given
       in  Table  \ref{tab2}. The $\mathrm{Al}-\mathrm{Ti}-\mathrm{O}$ phase diagram
        obtained from the computed formation energies is shown in Fig. \ref{f1}.

\begingroup \squeezetable
\begin{table}
      \caption{Calculated and experimental lattice parameters (in $\mathring{A}$) of
      materials  used
      as reference ones
      for building the $\mathrm{Al}-\mathrm{Ti}-\mathrm{O}$ phase diagram.} \label{tab1}
\begin{center}
     \begin{tabular} {|c|c|c|c|}\hline
      Crystal & Space group &  Lattice parameters& Lattice parameters \\
       &&(Theory)&(Experiment)\\ \hline
       Al&$Fm\bar{3}m$&$a=4.07$&$a=4.05$ \cite{18} \\
      $\mathrm{Al}_2\mathrm{O}_3$ &$R\bar{3}c$ & $a=4.86$, $c=13.19$&$a = 4.77$, $c = 13.00$
       \cite{19}\\
$\mathrm{TiO}_2$  &$P4_2/mnm$ & $a=4.69$, $c=2.99$ &  $a = 4.59$, $c = 2.96$\cite{20}\\
      $\mathrm{Ti}_2\mathrm{O}_3$ &$R\bar{3}c$& $a=5.14$, $c=14.10$ & $a = 5.16$, $c = 13.61$
      \cite{21}\\
            $\mathrm{TiO}$  &  $C2/m$ & $a=5.87$, $b=9.42$&$a=5.86$, $b=9.34$,\\&&$c=4.22$,
            $\gamma=107^{\circ}13'$&$c=4.14$, $\gamma=107^{\circ}32'$ \cite{22}\\
$\mathrm{Ti}_2\mathrm{O}$  &$P\bar{3}m1$&$a=3.01$, $c=4.86$& $a=2.92$, $c=4.71$\cite{23}
\\
$\mathrm{TiAl}$&$P4/mmm$ &$a=2.84$, $c=4.12$& $a=2.83$, $c=4.06$ \cite{18}\\
$\mathrm{TiAl}_2$&$I4_1/amd$ &$a=4.00$, $c=24.45$& $a=3.97$, $c=24.31$ \cite{24}\\
$\mathrm{TiAl}_{3}$&$I4/mmm$ &$a=3.87$, $c=8.69$& $a=3.84$, $c=8.58$ \cite{18}\\
 $\mathrm{Ti}$& $P6/mmm$      & $a=2.96$, $c=4.66$&   $a = 2.95$, $c =
      4.68$ \cite{18} \\
                \hline
      \end{tabular}
      \end{center}
\end{table}
\endgroup

\begin{table}
      \caption{Calculated formation energies $\Delta H_f$ per atom (in eV) for crystals listed
      in
      Table \ref{tab1}.} \label{tab2}
      \begin{center}
      \begin{tabular}{|c|c|c|c|c|c|c|c|c|}\hline
      Crystal   & $\mathrm{Al}_2\mathrm{O}_3$&
      $\mathrm{TiO}_2$&$\mathrm{Ti}_2\mathrm{O}_3$&$\mathrm{TiO}$& $\mathrm{Ti}_2\mathrm{O}$&
      $\mathrm{TiAl}$&$\mathrm{TiAl}_2$&$\mathrm{TiAl}_{3}$ \\\hline
      $\Delta H_f$ & -3.42 & -3.42  & -3.35 & -2.92  &-2.06 &-0.45 &-0.47 &-0.44\\
      \hline
      \end{tabular}
      \end{center}
      \end{table}

 We consider
the growth conditions in which the Al$_2$O$_3$ phase is in equilibrium with two other
phases. These conditions correspond to the points in the phase diagram specified in Table
\ref{tab3}. At the  point A the chemical potentials of Al, Ti and O atoms
        are related by the equations
\begin{eqnarray} \label{2}
            2\mu_{\mathrm{Al}} + 3\mu_\mathrm{O} &=& \mu_{\mathrm{Al}_2\mathrm{O}_3},\\
      \label{3}  \mu_{\mathrm{Ti}} + 2\mu_\mathrm{O} &=& \mu_{\mathrm{TiO}_2}, \\
      \label{4}  2 \mu_\mathrm{O} &=& \mu_{\mathrm{O}_2}.
      \end{eqnarray}
      The right-hand parts of Eqs. (\ref{2})-(\ref{4}) contain the energies
       (per formula unit) of the compounds  which are in equilibrium at the point A.
       Solving (\ref{2})-(\ref{4}) one gets
      the potentials
$\mu_{\mathrm{Al}}$, $\mu_{\mathrm{O}}$ and $\mu_{\mathrm{Ti}}$. Similar equations  can
be written for the points B, C etc.
 The oxygen potential
$\mu_{\mathrm{O}}$ (see Table \ref{tab3}) varies from the largest value at the point A
that corresponds to the oxidized limit to the  lowest value at the point F that
corresponds to the reduced limit. The value of  $\mu_{\mathrm{O}}$ given in Table
\ref{tab3} is counted from the chemical potential of an isolated oxygen atom.

\begin{table}
      \caption{Reference equilibrium points in the phase diagram Fig. \ref{f1}.} \label{tab3}
      \begin{center}
      \begin{tabular}{|c|c|c|}\hline
      Point label   &Compounds in equilibrium& $\mu_\mathrm{O}$(in eV)  \\
      &with  Al$_2$O$_3$ &\\\hline
      A & O$_2$, TiO$_2$ & -4.39 \\
 B & TiO$_2$, Ti$_2$O$_3$ & -8.18 \\
  C & Ti$_2$O$_3$, TiO & -9.46 \\
   D & TiO, TiAl$_2$& -9.77 \\
    E & TiAl$_2$,TiAl$_3$ & -9.87 \\
     F & TiAl$_3$, Al & -10.09 \\
      \hline
      \end{tabular}
      \end{center}
      \end{table}

      \section{Results and discussions}

      \subsection{Formation energies and binding energies}

The formation energy of an electrically neutral defect (Eq. (\ref{1}) with $q_i=0$) does
not depend on the Fermi energy $E_F$. The contribution of such defects into the free
energy and their equilibrium concentrations are independent of $E_F$ as well. On the
contrary, the formation energy of a charged defect (Eq. (\ref{1}) with $q_i\ne 0$)
depends on $E_F$. One can exclude $E_F$ from the problem considering electrically neutral
combinations of charged defects. The formation energy of any neutral combination is
independent of $E_F$. Taking into account overall charge neutrality of the crystal one
finds that the equilibrium concentrations of charged defects are independent of $E_F$ as
well. To compute these concentrations one should know only the formation energies of
electrically neutral combinations of charged defects.

Electrically neutral combinations can be chosen in an arbitrary way and a concrete choice
is just the matter of convention. The result of calculations of the equilibrium
concentrations will be the same for any choice.  Here we consider positively charged
defects in
 a combination with negatively charged
 $V_{\mathrm{Al}}^{3-}$ vacancies, and negatively charged defects, in a combination
with positively charged $\mathrm{Ti}^{4+}$ substitutional defects.

We restrict the analysis with complex defects formed by
substitutional $\mathrm{Ti}_\mathrm{Al}^{3+}$ and
$\mathrm{Ti}_\mathrm{Al}^{4+}$ ions, and $V_{\mathrm{Al}}^{3-}$
vacancies. Below we use the notations $\mathrm{Ti}^{3+}$ and
$\mathrm{Ti}^{4+}$ for Ti$_\mathrm{Al}$ substitutional defects. For
completeness we also take into account isolated interstitial ions (notated
as $\mathrm{Ti}^{3+}_i$
 and $\mathrm{Ti}^{4+}_i$), and  intrinsic defects  $V_{\mathrm{O}}^{2+}$,  $\mathrm{Al}_i^{3+}$
 and
$\mathrm{O}_i^{2-}$.  Intrinsic defects are considered in electrically neutral
combinations that correspond to the Schottky quintet
($3V_{\mathrm{O}}^{2+},2V_{\mathrm{Al}}^{3-}$), Al Frenkel
($\mathrm{Al}_i^{3+},V_{\mathrm{Al}}^{3-}$) pair and O Frenkel
 ($\mathrm{O}_i^{2-},V_{\mathrm{O}}^{2+}$) pair.

For each defect specie we compute the energy of the supercell with one isolated or
complex defect. A charge state of the defect is set as a  charge of the whole supercell.
The formation energy of an electrically neutral combination of defects
$\tilde{E}_\lambda$ is defined as the sum of the formation energies (\ref{1}) calculated
for the corresponding supercells and divided by the number of the supercells. The
quantity $\tilde{E}_\lambda$ yields the formation energy per defect and $\lambda$ stands
for a given electrically neutral combination.

We will show below that the concentration of complex defects can be
expressed through their binding energy. The binding energy
of a complex defect formed by $x$ $\mathrm{Ti}^{3+}$, $y$
$\mathrm{Ti}^{4+}$ and $z$ $V_{\mathrm{Al}}^{3-}$ vacancies is defined as
            \begin{equation}\label{eb}
        E_{i}^{(b)}=
        x E_3+y E_4 +z E_V-E_i,
       \end{equation}
where $E_{3}$, $E_4$ and $E_V$ are the formation energies (\ref{1}) of isolated
Ti$^{3+}$, Ti$^{4+}$  and $V_{\mathrm{Al}}^{3-}$ defects, correspondingly. Substituting
Eq. (\ref{1}) into the right-hand part of Eq. (\ref{eb}) we find that the chemical
potentials $\mu_{\mathrm{Al}}$, $\mu_{\mathrm{O}}$ and $\mu_\mathrm{Ti}$ are canceled.
Therefore, the binding energy is the same at different points of the ternary phase
diagram Fig. \ref{f1}. The binding energy can be expressed through the formation energies
of electrically neutral defects and the formation energies of electrically neutral
combinations of charged defects taken at the same equilibrium point.

\begin{figure}
     \center{\includegraphics[width=0.5\linewidth]{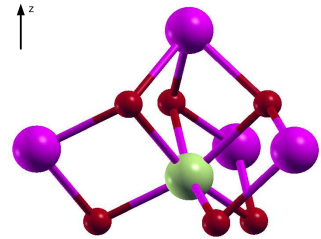}}
      \caption{Locations of four nearest neighbor Al atoms (large purple circles)
      around a given (central)  Al site
       (large green circle). Nearest neighbor O sites are shown by small red circles.
       The direction of the shortest $\mathrm{Al} -  \mathrm{Al}$
      link is shown by the arrow.}
        \label{f2}
      \end{figure}

The formation energy of  a complex defect and its binding energy depends on the distances
between simple defects which form the complex defect. In $\alpha$-Al$_2$O$_3$  every Al
atom has four nearest neighbor Al sites. These four sites form a tetrahedron (Fig.
\ref{f2}). The distances between a given Al site and four nearest neighbor Al sites are
almost the same (the link along the [0001] axis is shorter by 0.14 $\AA$). We consider
pairs where substitutional ions and Al vacancies are located at two nearest neighbor
sites (the central site and a tetrahedron apex shown in Fig. \ref{f2}). Four different
orientations of such pairs are possible. We obtain the same binding energy for three
orientations and a slightly different binding energy (the difference is about or less than
0.1 eV) for the fourth orientation (that corresponds to the shortest link). We neglect
this difference under the obtaining of the equilibrium concentrations of the defects.
 Triple defects under study are formed by single defects located at the
central site and at two tetrahedron apexes. We account two types of
$\mathrm{Ti}-\mathrm{Ti}-V$ complexes, one is with a Ti ion at the central site and the
other is with an Al vacancy at the central site. For the latter we use the notation
$\mathrm{Ti}-V-\mathrm{Ti}$. Triple defects can be in 6 different orientations. We
consider one quadruple defect specie. The quadruple defect is formed by a
$V_{\mathrm{Al}}^{3-}$ vacancy located at the central site surrounded with three  nearest
neighbor substitutional Ti ions.

The binding energies  are given in Table \ref{tab4}. One can see that all binding
energies except ones of $\mathrm{Ti}^{4+}-\mathrm{Ti}^{4+}$ and
$V_{\mathrm{Al}}^{3-}-V_{\mathrm{Al}}^{3-}$ pairs are positive. Negative binding energy
corresponds to the repulsion. Below we do not consider complex defects with negative
binding energy.

\begin{table}
      \caption{Computed binding energies of complex defects (in eV).} \label{tab4}
      \begin{center}
      \begin{tabular}{|c|c|}
      \hline
      Complex defect &  $E^{(b)}$   \\\hline
      $\mathrm{Ti}^{3+} -\mathrm{Ti}^{3+}$     & 1.36  \\
      $\mathrm{Ti}^{3+} - \mathrm{Ti}^{4+}$     & 0.70 \\
      $\mathrm{Ti}^{4+} - \mathrm{Ti}^{4+}$     & -0.11 \\
      $\mathrm{Ti}^{3+} - V_{\mathrm{Al}}^{3-}$ & 0.30\\
      $\mathrm{Ti}^{4+}-V_{\mathrm{Al}}^{3-}$ & 1.15   \\
 $V_{\mathrm{Al}}^{3-}-V_{\mathrm{Al}}^{3-}$ &  -3.25  \\
      $\mathrm{Ti}^{3+}-\mathrm{Ti}^{4+}-V_{\mathrm{Al}}^{3-}$  & 1.91    \\
$\mathrm{Ti}^{3+}-V_{\mathrm{Al}}^{3-}-\mathrm{Ti}^{4+}$   &  2.05    \\
      $\mathrm{Ti}^{4+}-\mathrm{Ti}^{4+}-V_{\mathrm{Al}}^{3-}$  & 1.30   \\
 $\mathrm{Ti}^{4+}-V_{\mathrm{Al}}^{3-}-\mathrm{Ti}^{4+}$   &  2.15 \\
 $\mathrm{Ti}^{3+}-\mathrm{Ti}^{3+}-V_{\mathrm{Al}}^{3-}$  & 2.16    \\
$\mathrm{Ti}^{3+}-V_{\mathrm{Al}}^{3-}-\mathrm{Ti}^{3+}$   & 1.19    \\
      $\mathrm{Ti}^{3+}-\mathrm{Ti}^{3+}-\mathrm{Ti}^{3+}$      & 2.03  \\
       $\mathrm{Ti}^{4+}-\mathrm{Ti}^{4+}-\mathrm{Ti}^{4+}-V_{\mathrm{Al}}^{3-}$  & 2.93   \\
            \hline
      \end{tabular}
      \end{center}
      \end{table}

\begin{figure}
      \begin{minipage}[h]{0.8\linewidth}
      \center{\includegraphics[width=1\linewidth]{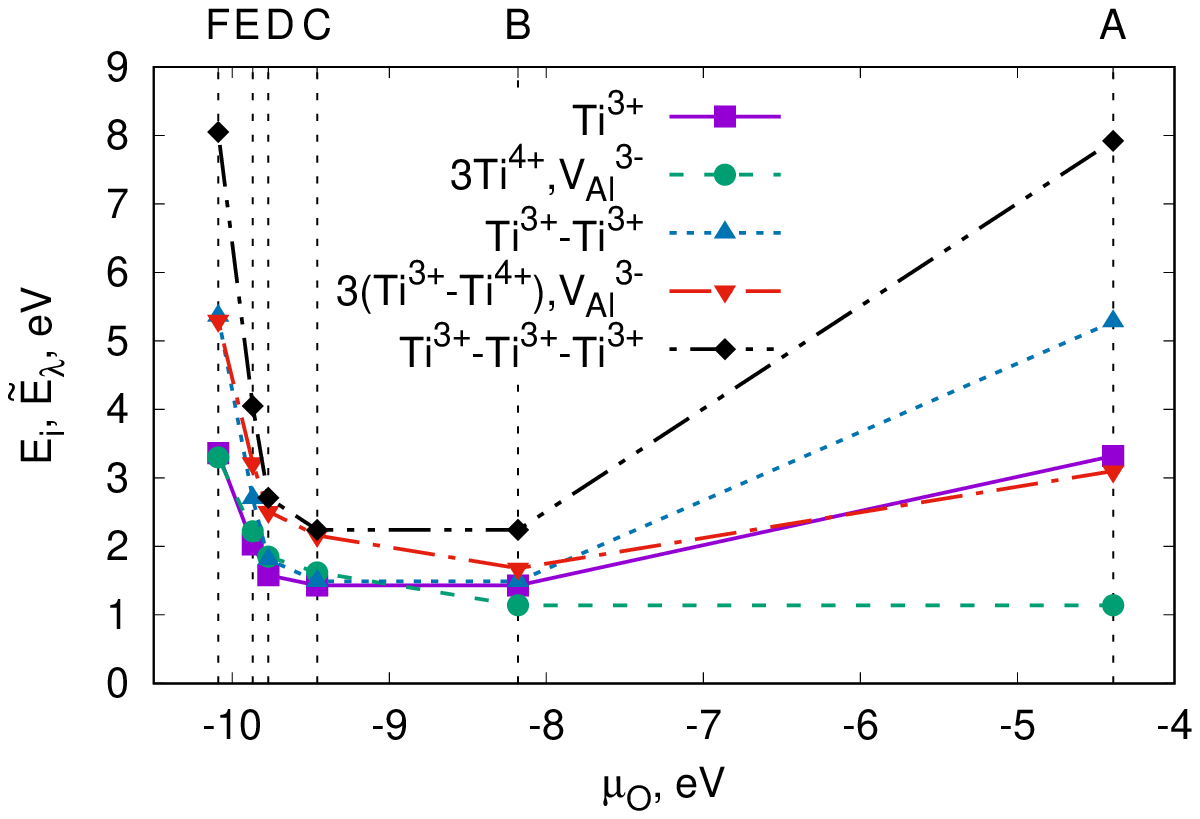}}
      \end{minipage}
      \begin{minipage}[h]{0.8\linewidth}
      \center{\includegraphics[width=1\linewidth]{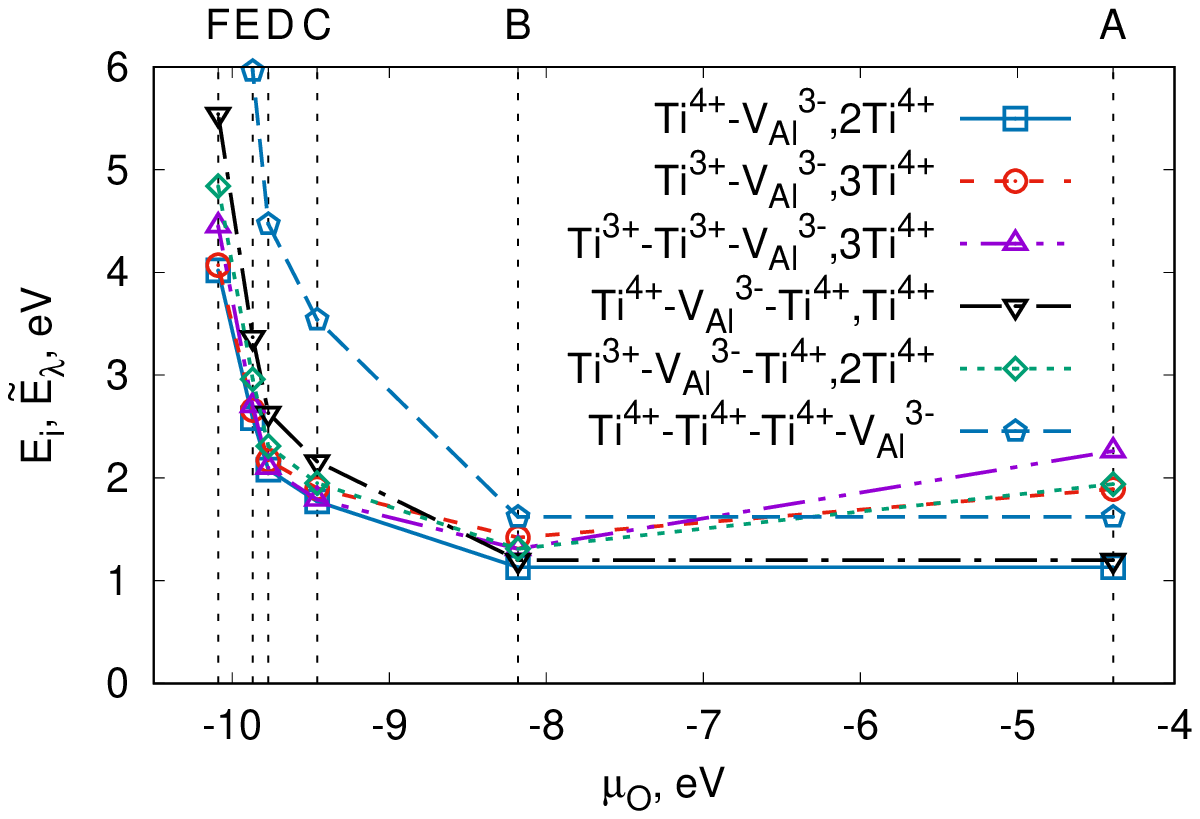}}
      \end{minipage}
      \begin{minipage}[h]{0.8\linewidth}
      \center{\includegraphics[width=1\linewidth]{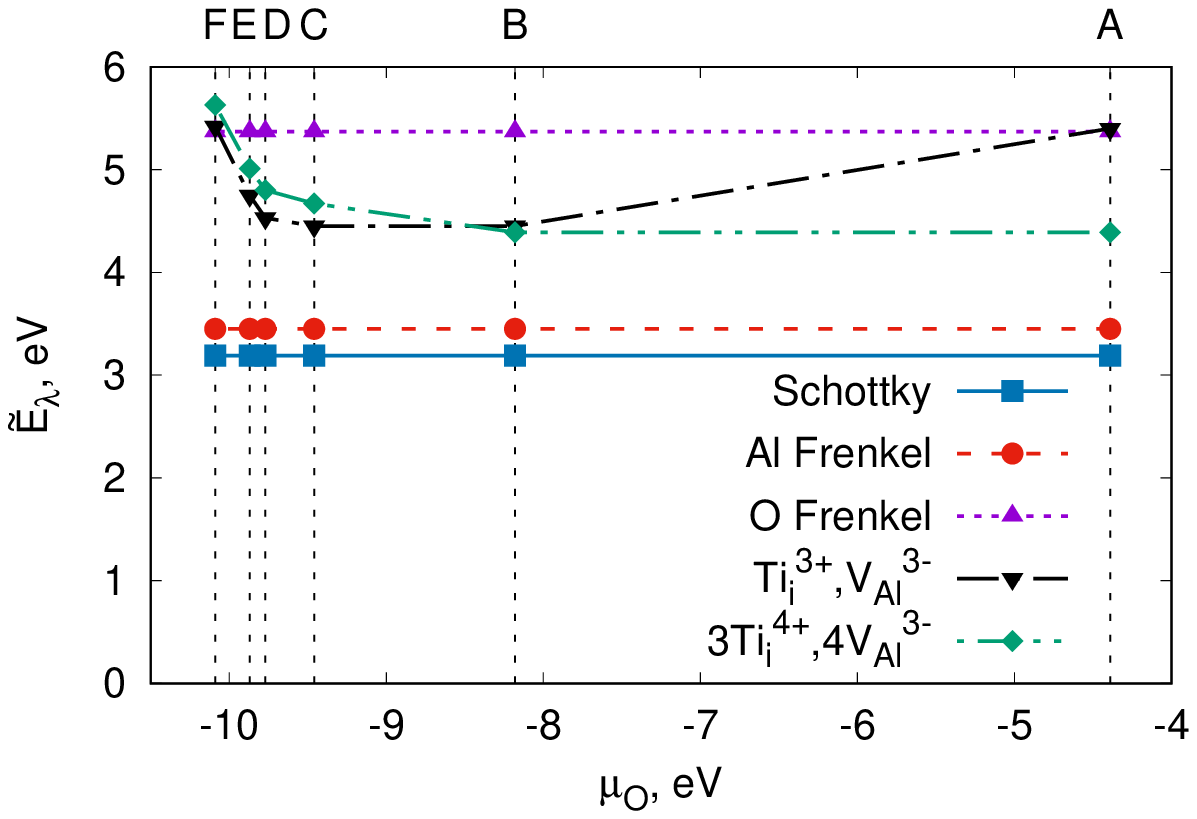}}
      \end{minipage}
\caption{Formation energies per defect of electrically neutral defects ($E_i$) and
electrically neutral combinations of charged defects ($\tilde{E}_\lambda$) at the
        equilibrium
         points A - F.
         Lines are guides to the eye. The value of the
          oxygen chemical potential  is shown in the abscissa axis.}
        \label{f3}
      \end{figure}

 The computed formation energies of  electrically neutral defects and of electrically neutral
 combinations
 of charged defects are presented  in  Fig. \ref{f3}.  The energies that correspond to
 energetically preferable orientations (configurations) of
complex defects  are displayed.

      \subsection{Equilibrium concentrations of defects}

       The free energy of defects in a crystal is given by the defect formation energies
       $E_i$,  the defect numbers $n_i$, and
       the configurational entropy (the logarithm of the number of ways
        $W_i$ to place $n_i$
defects in a crystal):
       \begin{equation}\label{6}
    F=\sum_i\left(E_i n_i - k_B T \ln W_i\right),
       \end{equation}
       where $k_B$ is the Boltzmann constant, and $T$ is the temperature.
       Since usually the  total number of  sites is much larger than the defect numbers each
       defect specie can be considered independently. Then
       \begin{equation}\label{7}
    W_i=\frac{N_i!}{(N_i-n_i)!n_i!},
       \end{equation}
       where $N_i$ is the number of ways to place one defect of a given specie into a
       crystal. Below we imply $N_i\gg n_i\gg 1$.

For isolated substitutional Ti defects and Al vacancies the quantity $N_i$ is the number
of Al atoms in the perfect crystal: $N_i=N_{\mathrm{Al}}$.  For complex defects  the
quantity $N_i$ accounts also different equivalent orientations of defects. It can be
expressed as  $N_i=a_i N_{\mathrm{Al}}$, where $a_{i}$ is a factor that depends on the
defect type. It is  equal to $a_i=2$ for $\mathrm{Ti}-\mathrm{Ti}$ pairs, $a_i=4$ for
$\mathrm{Ti}-V$ pairs and  $\mathrm{Ti}-\mathrm{Ti}-\mathrm{Ti}-V$ quadruples,  and
$a_i=6$
 for  the triples. For  Ti$_i$  the factor $a_i=1/2$ is the
 ratio of the number of empty O octahedral interstices to  $N_\mathrm{Al}$.

       The defect numbers $n_i$ are found from  the condition of minimum of the free
       energy. For neutral defect species the extremum condition yields
       \begin{equation}\label{8}
      \tilde{n}_i=\frac{n_i}{N_{i}}=\exp\left(-\frac{E_i}{k_B T}\right).
       \end{equation}

       The additional constraint for charged defects is
$\sum_i q_i n_i=0$, where $q_i$ is a defect charge in units of elementary charge. It is
convenient to separate the contribution of  Ti$^{4+}$ ($n_4$) and $V_\mathrm{Al}^{3-}$
($n_V$) in the charge neutrality constraint:
       \begin{equation}\label{9}
       n_4-3n_V + \sum_{i\ne 4,V} q_i n_i =0.
       \end{equation}
     Then, we exclude $n_V$ from the free energy (\ref{6}) and
       obtain the extremum conditions for the energy (\ref{6}) with respect to $n_4$
        and all other $n_{i}$:
        \begin{eqnarray}\label{10}
       E_4 + \frac{1}{3} E_V + k_B T
       \Big( \ln \tilde{n}_4 + \frac{1}{3} \ln \tilde{n}_V\Big)=0,
       \\\label{11}
       \frac{q_i}{3} E_V + E_i +
       k_B T \Big( \frac{q_i}{3} \ln \tilde{n}_V + \ln \tilde{n}_i \Big)=0,
       \end{eqnarray}
       where $\tilde{n}_i={n_i}/{N_{i}}$. One finds from Eqs. (\ref{10}) and (\ref{11})
       the relations
        \begin{equation}\label{12a}
       \tilde{n}_4^3 \tilde{n}_V  =\exp\Big(-\frac{4\tilde{E}_{4,V}}{k_B T}\Big),
       \end{equation} \begin{equation}\label{12}
       \tilde{n}_i = \tilde{n}_4^{q_i} \exp \Big( {-\frac{E_i - q_i E_4}{k_B T}} \Big),
\quad (i\ne 4,V),
       \end{equation}
    where $\tilde{E}_{4,V}=(3 E_4+E_V)/4$ is the formation energy per defect of a
       combination of three Ti$^{4+}$ and one $V_\mathrm{Al}^{3-}$.
       We note that Eq.~(\ref{12})
        describes the case of neutral defects  as well (at $q_i=0$ it reduces to Eq. (\ref{8})).

One can easily check that the energies $E_i - q_i E_4$ in Eq. (\ref{12}) are expressed
       through the formation energies $\tilde{E}_\lambda$. Respectively, the  equilibrium
       concentrations are determined by  $\tilde{E}_\lambda$ and do not depend on $E_F$.
  Substituting Eq. (\ref{12}) into Eqs.
        (\ref{9}) and taking (\ref{12a})
we have two algebraical equations for the variables $\tilde{n}_4$ and $\tilde{n}_V$.
These equations have a unique positive real-valued solution. From this solution and Eq.
(\ref{12}) we obtain  the concentrations of all considered defect species.

       Eqs. (\ref{12a}) and ({\ref{12}) allow us to express the concentrations of
        complex defects through the concentrations of isolated defects.
       The equilibrium concentration of complex defects composed of $r_3$
$\mathrm{Ti}^{3+}$ ions, $r_4$  $\mathrm{Ti}^{4+}$ ions and $r_V$ $V_\mathrm{Al}^{3-}$
vacancies is equal to
       \begin{equation}\label{25}
      \tilde{n}_i=
      (\tilde{n}_3)^{r_3}
      (\tilde{n}_4)^{r_4}
      (\tilde{n}_V)^{r_V}
      \exp\left(\frac{E_i^{(b)}}{k_B T}\right),
       \end{equation}
where $E_i^{(b)}$ is the binding energy (\ref{eb}).
 Eq. (\ref{25}) can be applied
instead of Eq. (\ref{12}) to calculate equilibrium concentrations of complex defects. Eq.
(\ref{25}) can be also useful if the concentrations of isolated defects are known, for
instance, from experimental data.

 In the charge neutrality
       equation (\ref{9}) it is enough to take into account Ti$^{4+}$ and
       $V_\mathrm{Al}^{3-}$ isolated defects, and some other defect species with the
       smallest  formation energies. Then solving the obtained system for $\tilde{n}_4$
and $\tilde{n}_V$ and substituting the answer into Eq. (\ref{12}) or Eq. (\ref{25}) one
finds the concentrations of the rest defect species.
              Such an approximation
       is valid if the overall concentration of species neglected in the charge
        neutrality equation is small in comparison with
         the concentration of species accounted in it.
         Relative error
        in determining the concentrations is proportional to that small parameter.

             Below we exclude the
point F from the consideration and account Ti$^{4+}$,
       $V_\mathrm{Al}^{3-}$, $\mathrm{Ti}^{4+}-V_\mathrm{Al}^{3-}$ and
$\mathrm{Ti}^{4+}-V_\mathrm{Al}^{3-}-\mathrm{Ti}^{4+}$ defects in the charge neutrality
equation. Using Eq. (\ref{25}) we obtain
\begin{equation}\label{9a}
\tilde{ n}_4-3\tilde{n}_V  - 8 \tilde{n}_{4} \tilde{n}_{V}e^\frac{E_{4V}^{(b)}}{k_B T} -6
\tilde{n}_4^2 \tilde{n}_V e^\frac{E_{4V4}^{(b)}}{k_B T} =0,
\end{equation}
where $E_{4V}^{(b)}$ and $E_{4V4}^{(b)}$ are the binding energies of
$\mathrm{Ti}^{4+}-V_\mathrm{Al}^{3-}$ pairs and
$\mathrm{Ti}^{4+}-V_\mathrm{Al}^{3-}-\mathrm{Ti}^{4+}$ triples, correspondingly. The
solution  of Eqs. (\ref{12a}) and (\ref{9a}) can be presented in the form
\begin{equation}\label{9b}
\tilde{n}_4 = z\tilde{n}_4^{(0)},
\end{equation}
where $\tilde{n}_4^{(0)}=3^{1/4}\exp(-\tilde{E}_{4,V}/k_B T)$ is the approximate solution
 obtained under accounting of only Ti$^{4+}$ and $V_\mathrm{Al}^{3-}$  in the
  charge neutrality equation (\ref{9a}).
  The factor
$z$ satisfies the equation
\begin{equation}\label{9c}
    z^4-\alpha z - \beta z^2 = 1.
\end{equation}
The coefficients $\alpha$ and $\beta$ in Eq. (\ref{9c}) are equal to
\begin{equation}\label{9d}
    \alpha
    =\frac{8}{3^{3/4}}e^{-\frac{3\left(\tilde{E}_{4V,4}-\tilde{E}_{4,V}\right)}{k_B
    T}},
\end{equation}
\begin{equation}\label{9d1}
    \beta
    =\frac{6}{\sqrt{3}}e^{-\frac{2\left(\tilde{E}_{4V4,4}-\tilde{E}_{4,V}\right)}{k_B
    T}},
\end{equation}
where $\tilde{E}_{4V,4}$ and $\tilde{E}_{4V4,4}$ are the formation energies (per defect)
of the corresponding electrically neutral combinations (one
$\mathrm{Ti}^{4+}-V_\mathrm{Al}^{3-}$ pair plus two isolated $\mathrm{Ti}^{4+}$ and one
$\mathrm{Ti}^{4+}-V_\mathrm{Al}^{3-}-\mathrm{Ti}^{4+}$ triple plus one isolated
$\mathrm{Ti}^{4+}$, respectively).

The factor $z$ is defined as a positive real-valued root of Eq. (\ref{9c}).
 For $\tilde{E}_{4V,4}-\tilde{E}_{4,V} \gg k_B T$ and $\tilde{E}_{4V4,4}-\tilde{E}_{4,V}
 \gg k_B T$ the coefficients $\alpha$
and $\beta$ approach zero and $\tilde{n}_4$ coincides with $\tilde{n}_4^{(0)}$.

        The concentrations of other charged defects are calculated
         from Eq. (\ref{12}). Substituting Eq. (\ref{9b}) into Eq. (\ref{12})
         we find that these quantities are
         expressed through
         the formation energies $\tilde{E}_\lambda$.
       In particular, for positively charged defects ($q_i>0$)
        \begin{eqnarray}
            \label{18}
        {\tilde{n}_{i}}&\approx& {\tilde{n}_4} (\sqrt[4]{3}z)^{q_i-1}
        e^{-\left(1 + \frac{q_{i}}{3}\right)\frac{\tilde{E}_{i,V}-
        \tilde{E}_{4,V}}{ k_B T}},
 \end{eqnarray}
where the index $\lambda=i,V$ stands for a combination of
     three   positively charged defects and $q_i$  $V_\mathrm{Al}^{3-}$ vacancies.
For negatively charged defects ($q_i<0$)
\begin{eqnarray}
         \label{18a}
          \tilde{n}_{i} &\approx&
        \frac{\tilde{n}_4}{(\sqrt[4]{3}z)^{1+|q_{i^-}|}}
        e^{-(1+|q_{i}|)\frac{\tilde{E}_{i,4}- \tilde{E}_{4,V}}{k_B
        T}},
    \end{eqnarray}
    where  $\lambda=i,4$ stands for a combination
    of one negatively charged defect and $|q_i|$ substitutional Ti$^{4+}$.

 The
concentrations of  intrinsic defects obtained in a similar way are equal to
\begin{eqnarray} \label{18b}
  \tilde{n}_{V_\mathrm{Al}^{3-}} &\approx& \frac{\tilde{n}_4}{3z^4}, \\ \label{18c}
 \tilde{n}_{V_\mathrm{O}^{2+}} &\approx& {\tilde{n}_4}\sqrt[4]{3}z
 e^{-\frac{5}{3}\frac{\tilde{E}_\mathrm{Sch}-\tilde{E}_{4,V}}{k_B
        T}},\\ \label{18d}
\tilde{n}_{\mathrm{Al}_i^{3+}} &\approx& \tilde{n}_4(\sqrt[4]{3}z)^2
e^{-\frac{2(\tilde{E}_{\mathrm{Al}\, \mathrm{Frenkel}}-\tilde{E}_{4,V})}{k_B
        T}},\\
  \label{18e}
  \tilde{n}_{\mathrm{O}_i^{2-}} &\approx& \frac{\tilde{n}_4}{(\sqrt[4]{3}z)^3}
e^{-\frac{2\tilde{E}_{\mathrm{O}\, \mathrm{Frenkel}}-
\frac{5}{3}\tilde{E}_{\mathrm{Sch}}-\frac{1}{3}\tilde{E}_{4,V}}{k_B
        T}},
\end{eqnarray}
where $\tilde{E}_{\mathrm{Frenkel}\, \mathrm{Al}(\mathrm{O})}$ and
$\tilde{E}_{\mathrm{Sch}}$ are the formation energies per one defect site for  O(Al)
Frenkel pair  and for the Schottky quintet, respectively. Note that Eqs.
(\ref{18c})-(\ref{18e}) are applicable if the concentrations of the corresponding defects
are small compared to the concentration of substitutional Ti$^{4+}$ ions. The latter is
provided by the smallness of the formation energy $\tilde{E}_{4,V}$ in comparison with
the Schottky and Frenkel defect formation energies. We emphasize that Eqs.
(\ref{18b})-(\ref{18e}) cannot be applied to pure Al$_2$O$_3$. Considering pure crystals
one should account only intrinsic defects in the charge neutrality equation. It results
in different from (\ref{18b})-(\ref{18e}) equilibrium concentrations.

It is instructive to find   distribution of Ti impurities between different defect species.
 The partial concentration of Ti that corresponds to the $i$-th defect specie is given by the
  expression
    \begin{equation}\label{20}
    w_i=\tilde{n}_i a_i k_i\frac{2
    m_{\mathrm{Ti}}}{m_{\mathrm{Al}_2\mathrm{O}_3}}\cdot 100\% ,
    \end{equation}
    where    $m_{\mathrm{Ti}}$ and $m_{\mathrm{Al}_2\mathrm{O}_3}$ are the molecular
    masses of Ti and $\mathrm{Al}_2\mathrm{O}_3$,
     $k_i$ is the number of Ti atoms in a given defect and $a_i$ is the orientation
     factor defined above. The quantities $w_i$ are  the partial concentrations   in percent
     by mass  (wt\%).
Calculated $w_i$ at the temperature $T=1600$ K  are shown in Fig. \ref{f4}.
      The sum $w_{\mathrm{Ti}}=\sum_i w_i$ yields the total equilibrium concentration
     of Ti. This quantity is also shown in  Fig. \ref{f4}. One finds from Fig. \ref{f4}
     that the overall concentration of charged defects neglected in Eq. (\ref{9a}) is
     about or less than one hundredth of Ti$^{4+}$ concentration.      The concentrations of
     $\mathrm{Ti}^{3+}_i$ and $\mathrm{Ti}^{4+}_i$ interstitial defects are too small  and out of
      range of $w_i$  in Fig. \ref{f4}.

One can see from Fig.~\ref{f4} that in the oxidized limit (point A) Ti is mainly in the
form of isolated substitutional $\mathrm{Ti}^{4+}$ ions, while in the reduced limit
(point E) it is in the form of isolated substitutional $\mathrm{Ti}^{3+}$. In the
oxidized (point A) and intermediate (point B) conditions a noticeable part of
$\mathrm{Ti}^{4+}$ ions  form pairs, triples or quadruples with $V_{\mathrm{Al}}^{3-}$
vacancies. In the intermediate conditions (points B and C) most of $\mathrm{Ti}^{3+}$
ions bind in $\mathrm{Ti}^{3+}-\mathrm{Ti}^{3+}$ pairs. The concentration of
$\mathrm{Ti}^{3+}-\mathrm{Ti}^{4+}$ pairs  reaches its maximum in the intermediate
conditions (point B), but it remains much smaller than the concentration of isolated
$\mathrm{Ti}^{3+}$. At the point B  most of $\mathrm{Ti}^{3+}-\mathrm{Ti}^{4+}$ pairs
bind in triples with $V_{\mathrm{Al}}^{3-}$ vacancies.  The largest relative
concentration (about $10^{-2}$ of the total Ti concentration) of triples
$\mathrm{Ti}^{3+}-\mathrm{Ti}^{3+}-\mathrm{Ti}^{3+}$ is reached at the point C.
 Over the entire range of $\mu_\mathrm{O}$ the amount of Ti distributed between
 other
  defects is smaller than $10^{-2}$ of the total Ti amount.

The computed concentrations of intrinsic defects are shown in Fig. \ref{f5}. One can see
that the concentrations of native defects excluding $V_{\mathrm{Al}}^{3-}$  are rather
small, and therefore one can omit their contribution into the charge neutrality equation
(\ref{9}).

    \begin{figure}[h!]
 \center{\includegraphics[width=1\linewidth]{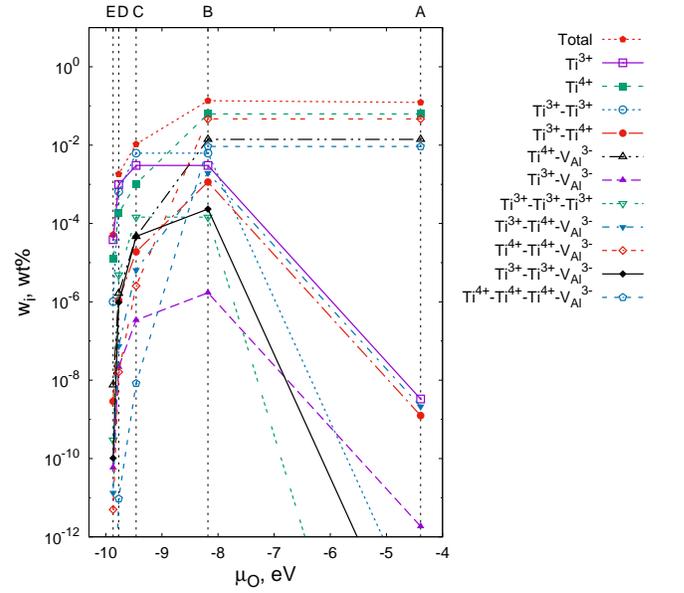}}
          \caption{Thermodynamically equilibrium total concentration of
          Ti in percent by mass and its distribution between defect species at the equilibrium
           points A - E at the temperature $T=1600 K$. Lines are guides to the eye.}\label{f4}
    \end{figure}

\begin{figure}[h!]
      \center{\includegraphics[width=1\linewidth]{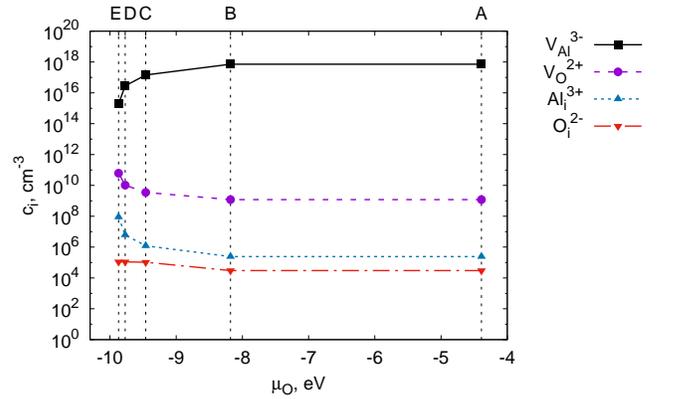}}
              \caption{Thermodynamically equilibrium concentrations of
          native defects at the temperature $T=1600 K$.}\label{f5}
    \end{figure}

The results presented in Figs. \ref{f4} and \ref{f5} have been obtained neglecting
temperature effects on the chemical potential of the precursors.  It is connected with
that  DFT is the zero-temperature technique. It was proposed in \cite{t1,t2} to account
temperature effects  within the approach that combines DFT and classical thermodynamics.
In \cite{t1,t2} the  approach was used for the obtaining of $(T,p)$ phase diagrams of
surface structures. Similar approach was implemented in \cite{11} for the calculation of
the formation energies of native defects in Al$_2$O$_3$. Below we account  temperature
corrections to the defect formation energies in Ti-doped Al$_2$O$_3$ and consider how
these corrections influence the equilibrium concentrations of the defects.

The temperature correction to the chemical potential of a precursor is given by the
equation \cite{t1,t2}
\begin{equation}\label{en1}
\Delta\mu(T,p_0)=\Delta H(T,p_0)-\Delta H(0,p_0)-T S(T,p_0),
\end{equation}
where $\Delta H(T,p_0)$ is the difference of enthalpy of the precursor at temperature $T$
and at the reference temperature $T_r=298.15$ K, and $S(T,p_0)$ is entropy at the
standard pressure $p_0=0.1$ MPa. The quantities $\Delta H(0,p_0)$, $\Delta H(T,p_0)$ and
$S(T,p_0)$ can be taken from the thermochemical tables \cite{tt}. The obtained
$\Delta\mu$ at $T=1600$ K are given in Table \ref{tab5}.

\begingroup \squeezetable
\begin{table}
      \caption{Temperature correction to the chemical potentials of precursors
       per formula unit (in eV) at $T=1600$ K.}
       \label{tab5}
      \begin{center}
      \begin{tabular}{|c|c|c|c|c|c|c|c|}\hline
      Precursor   & Al&
      Ti&$\mathrm{Al}_2\mathrm{O}_3$&$\mathrm{TiO}_2$& $\mathrm{Ti}_2\mathrm{O}_3$&
      $\mathrm{TiO}$&$\mathrm{O}_2$ \\\hline
      $\Delta\mu(1600,p_0)$ & -0.901 & -0.883  & -2.278 & -1.713  &-3.020 &-1.232 &-3.770\\
      \hline
      \end{tabular}
      \end{center}
      \end{table}
\endgroup

Using the data of Tables \ref{tab2} and \ref{tab5} we plot the Al-Ti-O phase diagram Fig.
\ref{f6} that accounts the temperature corrections to the formation energies (\ref{5}) at
$T=1600$ K. The equilibrium points A, B and C in Fig. \ref{f6} are the same as in the
diagram Fig. \ref{f1}. The phases TiAl$_n$ ($n=1,2,3$) are not included in the diagram
Fig. \ref{f6} due to  lack of thermochemical data for these compounds in \cite{tt}. The
points $D$, $E$ and $F$ are replaced with the point F$'$ that corresponds to the
equilibrium between Al$_2$O$_3$, TiO and the liquid phase of Al. Taking into account the
correction (\ref{en1}) we obtain the temperature corrections to the chemical potentials
of atoms $\Delta\mu_{X}(T)$ ($X=\mathrm{Al}, \mathrm{Ti}, \mathrm{O}$) at the equilibrium
points A, B, C and F$'$.

\begin{figure}[h]
     \center{\includegraphics[width=0.4\textwidth]{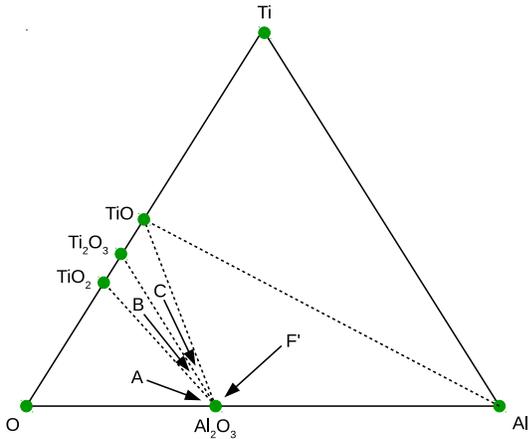}}
     \caption{Phase diagram of the Al-Ti-O ternary system at $T=1600$ K obtained in the approach
     that combines
DFT and classical thermodynamics.}\label{f6}
     \end{figure}

The defect formation energy (\ref{1}) contains the difference of energies of a defect and
the perfect supercell $E_{def-perf}=E_{def}-E_{perf}$. The DFT approach yields this
difference   at $T=0$ K. In the
 combined  approach  one should
 account not only the temperature corrections to the chemical potentials of atoms
  but  a temperature correction to the difference $E_{def-perf}$, as well.
  For the supercell with substitutional Ti atoms we evaluate this
  correction as the difference of $\Delta\mu_{\mathrm{Ti}}(T)$ in Ti$_2$O$_3$ and
  $\Delta\mu_{\mathrm{Al}}(T)$ in Al$_2$O$_3$. These compounds have the same crystal structure
  and the
  difference in $\Delta H$ and $S$ is caused in the main part by the
  replacement of Ti with Al. It yields $\Delta
  E_{def-perf}=-0.371$ eV at $T=1600$ K per one substitutional Ti atom.
  For the supercell with one Al vacancy the temperature correction
   can be evaluated as $\Delta \mu(T,p_0)$ for Ti crystal (see Table \ref{tab5}) minus $\Delta
  E_{def-perf}$ for the supercell with one substitutional Ti ion   (we use
  Ti crystal as a reference compound since Al is in a liquid phase at $T=1600$ K). It
   yields $\Delta
  E_{def-perf}=-0.513$ eV at $T=1600$ K per one vacancy.
  Taking into account the temperature corrections  $\Delta\mu_\mathrm{Al}$,
  $\Delta\mu_\mathrm{Ti}$, $\Delta\mu_\mathrm{O}$ and $\Delta
  E_{def-perf}$ in Eq. (\ref{1})
  we calculate  the defect formation energies at $T=1600$ K. Using these energies we obtain from
  Eqs. (\ref{9})-(\ref{12a})  the
  concentrations of the defects. The result is presented in Fig. \ref{f7}.
  One can see that the  concentrations obtained from the
   formation energies in which the temperature corrections are neglected  (Fig. \ref{f4})
   and from the formation energies that account the temperature corrections (Fig. \ref{f7})
    demonstrate
   basically the same behavior  under variation
   of the oxygen chemical potential. There are some minor differences, namely, in the oxidized
    conditions  the calculations that
    account the temperature corrections yield higher concentrations of
     Ti$^{3+}$ ions in comparison with
    the calculations
    where such  corrections are neglected. In the reduces conditions
     the concentration of
     Ti$^{4+}$ ions is underestimated  if the temperature corrections are neglected.
    In the intermediate conditions the temperature corrections to the defect formation
    energies are rather small and result only in unessential changes of equilibrium
    concentrations of the defects.

  \begin{figure}[h!]
 \center{\includegraphics[width=1\linewidth]{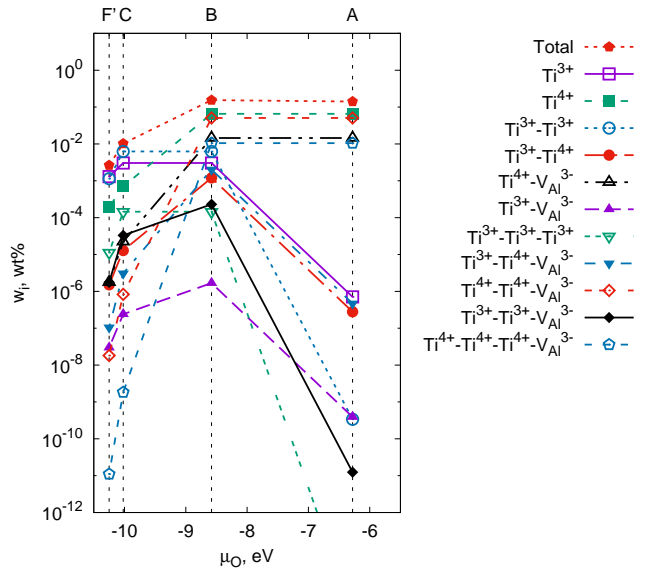}}
          \caption{Thermodynamically equilibrium total and partial concentrations of
          Ti  at the temperature $T=1600 K$ obtained in the
          approach that combines
DFT and classical thermodynamics.}\label{f7}
    \end{figure}

        \subsection{Equilibrium concentration
        of defects in Ti-deficient conditions }

   Let us now consider a  situation where the total concentration of Ti
      is less than the equilibrium one. To obtain partial concentrations of defects
     one should minimize the
 free energy (\ref{6}) with the additional constraint
    \begin{equation}\label{21}
    \sum_i k_i n_i=n_\mathrm{Ti},
    \end{equation}
    where $n_\mathrm{Ti}$ is the  total number of Ti atoms in the crystals.
 In this case we exclude
 $n_3$ (the number of Ti$^{3+}$) and $n_V$
  from the free energy (\ref{6}) and
       find the extremum condition for the free energy (\ref{6}) with respect to $n_4$
        and all other $n_{i}$.  We arrive at the  equations
        \begin{eqnarray}\label{50}
       E_4 - E_3 + \frac{1}{3} E_V \cr + k_B T
       \Big( \ln \tilde{n}_4  + \frac{1}{3} \ln \tilde{n}_V-\ln \tilde{n}_3\Big)=0,
       \\\label{51}
       \frac{q_i}{3} E_V + E_i -k_i E_3 +\cr
       k_B T \Big( \frac{q_i}{3} \ln \tilde{n}_V + \ln \tilde{n}_i
       -k_i \ln \tilde{n}_3\Big)=0.
       \end{eqnarray}

      From Eqs. (\ref{50}) and (\ref{51}) we obtain the following relations for the concentrations
    \begin{equation}\label{27}
        \tilde{n}_4^3 \tilde{n}_V
        =\tilde{n}_3^3\exp\Big(\frac{3 E_3-4\tilde{E}_4
        }{k_B T}\Big),
          \end{equation}
        \begin{equation}\label{28}
        \tilde{n}_i = \tilde{n}_4^{q_i}
        \tilde{n}_3^{k_i-q_i}
        \exp \Big( {\frac{(k_i-q_i)E_3-E_i + q_i E_4}{k_B T}} \Big).
        \end{equation}
   Using Eqs.~(\ref{27}) and (\ref{28}) we find that the
    relation between the  concentrations
   of isolated and complex defects is exactly the same as above (Eq. (\ref{25})).
   It is remarkable that the restriction
(\ref{21}) does not change this relation. Substituting the relation (\ref{25}) into  Eqs.
(\ref{9}), (\ref{21}) and solving them together with Eq. (\ref{27}) we obtain
$\tilde{n}_3$, $\tilde{n}_4$ and $\tilde{n}_V$.

    Let us restrict ourselves with seven defect species: three isolated defects (Ti$^{3+}$,
     Ti$^{4+}$ and $V_{\mathrm{Al}}^{3-}$)
    and four complex defects ($\mathrm{Ti}^{3+}-\mathrm{Ti}^{3+}$ pairs,
     $\mathrm{Ti}^{4+}-V_{\mathrm{Al}}^{3-}$ pairs,
       $\mathrm{Ti}^{4+}-V_{\mathrm{Al}}^{3-}-\mathrm{Ti}^{4+}$
       triples
       and  $\mathrm{Ti}^{4+}-\mathrm{Ti}^{4+}-\mathrm{Ti}^{4+}-V_{\mathrm{Al}}^{3-}$
       quadruples). Within this simplification Eqs. (\ref{9}) and
(\ref{21}) are reduced to the following ones
        \begin{eqnarray}
    \label{40}
      \tilde{n}_4 - 3 \tilde{n}_V - 8  \tilde{n}_{4} \tilde{n}_{V}
      e^\frac{E_{4V}^{(b)}}{k_B T}- 6 \tilde{n}_{4}^2
       \tilde{n}_{V} e^\frac{E_{4V4}^{(b)}}{k_B T}=0,
       \\ \label{41}
      \tilde{n}_3 + \tilde{n}_4 + 4 \tilde{n}_{3}^2 e^\frac{E_{33}^{(b)}}{k_B T}
      + 4 \tilde{n}_{4} \tilde{n}_{V} e^\frac{E_{4V}^{(b)}}{k_B T}
      \cr + 12 \tilde{n}_{4}^2 \tilde{n}_{V} e^\frac{E_{4V4}^{(b)}}{k_B T}
       + 12 \tilde{n}_{4}^3 \tilde{n}_{V} e^\frac{E_{4V}^{(b)}}{k_B T}=
      \frac{n_\mathrm{Ti}}{N_\mathrm{Al}},
    \end{eqnarray}
where $E_{33}^{(b)}$ is the binding energy of $\mathrm{Ti}^{3+}-\mathrm{Ti}^{3+}$ pairs.
Solving the system (\ref{27}), (\ref{40}), (\ref{41}) we obtain   the concentrations
$\tilde{n}_{i}$. In Fig. \ref{f8} the distribution of Ti (in wt\%) between different
defect species at the points $B$ and $C$ of the phase diagram is shown. Since in the
intermediate conditions the temperature corrections to the defect formation energies are
small we neglect them in the calculations.
 One can
see that the deficit of Ti results in two effects. First, the relative fraction of complex
defects decreases. Second, the ratio of isolated  $\mathrm{Ti}^{4+}$ ions to
$\mathrm{Ti}^{3+}$ ions increases.

\begin{figure}[h!]
\begin{minipage}[h]{1.0\linewidth}
\center{\includegraphics[width=1\linewidth]{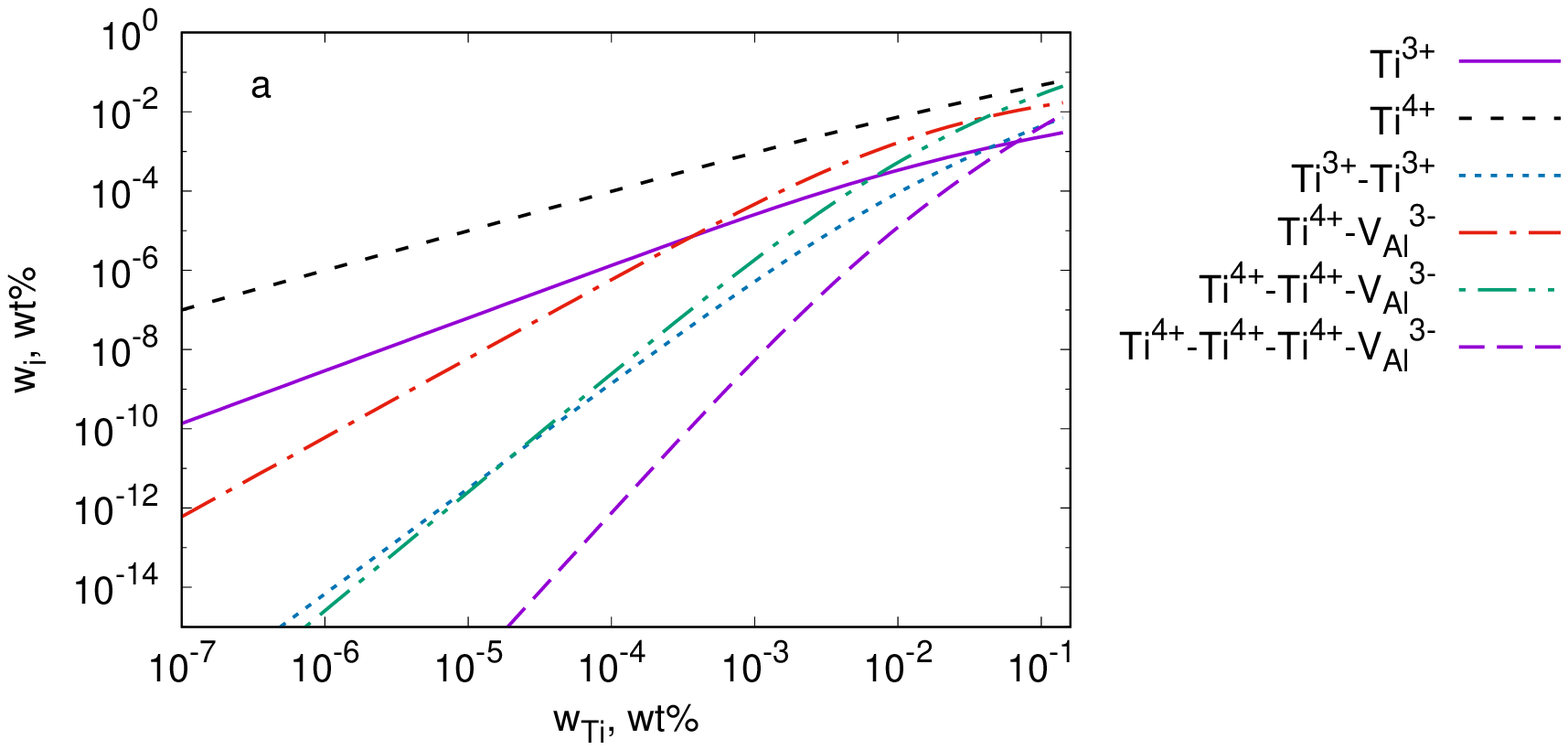}}   \\
\end{minipage}
\begin{minipage}[h]{1.0\linewidth}
\center{\includegraphics[width=1\linewidth]{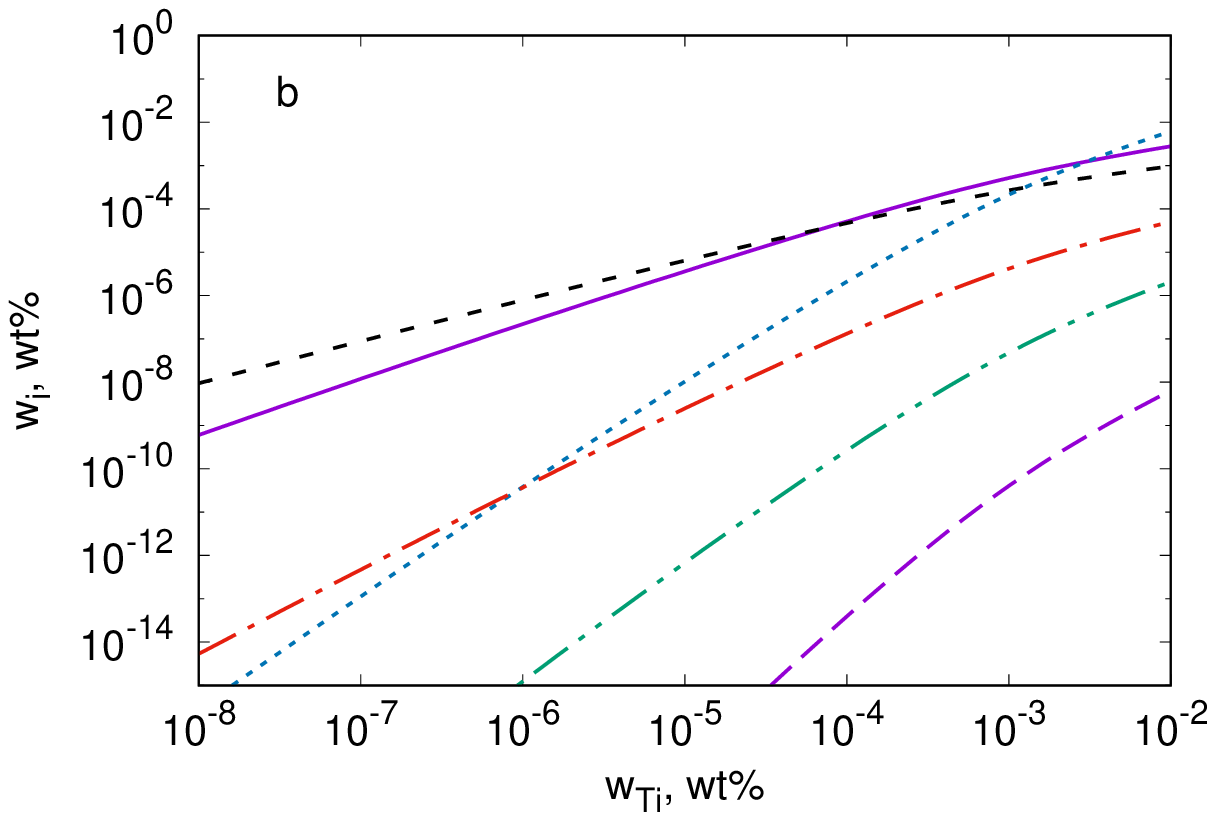}} \\
\end{minipage}
\caption{Contributions of  different defect species into the  total Ti concentration
$w_\mathrm{Ti}$ in the Ti-deficient conditions at the equilibrium points B (a) and C (b)
of the phase diagram.} \label{f8}
\end{figure}

\subsection{Ti impurity levels in the band structure}

The presence of Ti in $\alpha$-$\mathrm{Al}_2\mathrm{O}_3$ results in an appearance of
additional levels in the band gap. To obtain these levels we compute the band structures
of $\alpha$-$\mathrm{Al}_2\mathrm{O}_3$ with one defect  in the supercell. A 120-atom
supercell  with one, two and three Ti atoms contains 1.9 wt\%, 3.8 wt\% and 5.7 wt\%  of
Ti, correspondingly. It is much larger than the equilibrium concentration of Ti
calculated above, but since our band structure calculations yield very narrow impurity
 bands (see below) such calculations describe adequately impurity levels at low Ti
concentration.

In Fig \ref{f9}a the calculated band structure of the perfect
$\alpha$-$\mathrm{Al}_2\mathrm{O}_3$ is shown. The Brillouin zone corresponds to the
120-atom
 supercell.  The calculated  band gap is $E_g=4.91$ eV. Underestimation of $E_g$
(experimental value is 8.7 eV) is a known band gap prediction problem of the standard DFT
methods based on the PBE functional \cite{25} as well as on the local-density
approximation (see, for instance \cite{26}). In \cite{26} it was proposed to use the
modified Becke-Johnson (MBJ) potential for accurate calculations of the band gap. The MBJ
potential is a modified version of the original Becke-Johnson (BJ) exchange potential
\cite{bj}. The MBJ potential contains the Becke-Roussel (BR) potential \cite{br} instead
of the Slater exchange potential. The BR potential and the MBJ potentials are quasilocal
ones: they are determined by the electron density and its first and second spatial
derivatives. In \cite{27} the MBJ potential was employed to study the structural,
electronic, and optical properties of doped $\alpha$-Al$_2$O$_3$.
 In
\cite{27} the band gap of pure $\alpha$-$\mathrm{Al}_2\mathrm{O}_3$ was calculated by the
full potential linear augmented plane wave  method with the PBE and MBJ potentials. The
PBE calculations yields $E_g=6.5$ eV, but  the result of MBJ calculations is very close
($E_g=8.5$ eV) to the experimental value. The authors of \cite{27} considered impurity
levels of yttrium, scandium, zirconium and niobium doped
$\alpha$-$\mathrm{Al}_2\mathrm{O}_3$ crystals. They found that the same impurity induced
peaks in the density of states appear in the PBE and mBJ calculations. These peaks have
the same orbital decomposition and the relative positions within the gap, although their
width and absolute positions are sensitive to the potential used. The authors of
\cite{27} have concluded that the PBE and mBJ calculations of impurity levels produce
qualitatively similar results, which can be interpreted in the same way. Basing on this
conclusion we expect that the method implemented in SIESTA code allows to analyze
adequately impurity levels in Ti-doped $\alpha$-$\mathrm{Al}_2\mathrm{O}_3$ with complex
defects.

\begin{figure}
\begin{minipage}{0.7\linewidth}
\center{\includegraphics[width=1\linewidth]{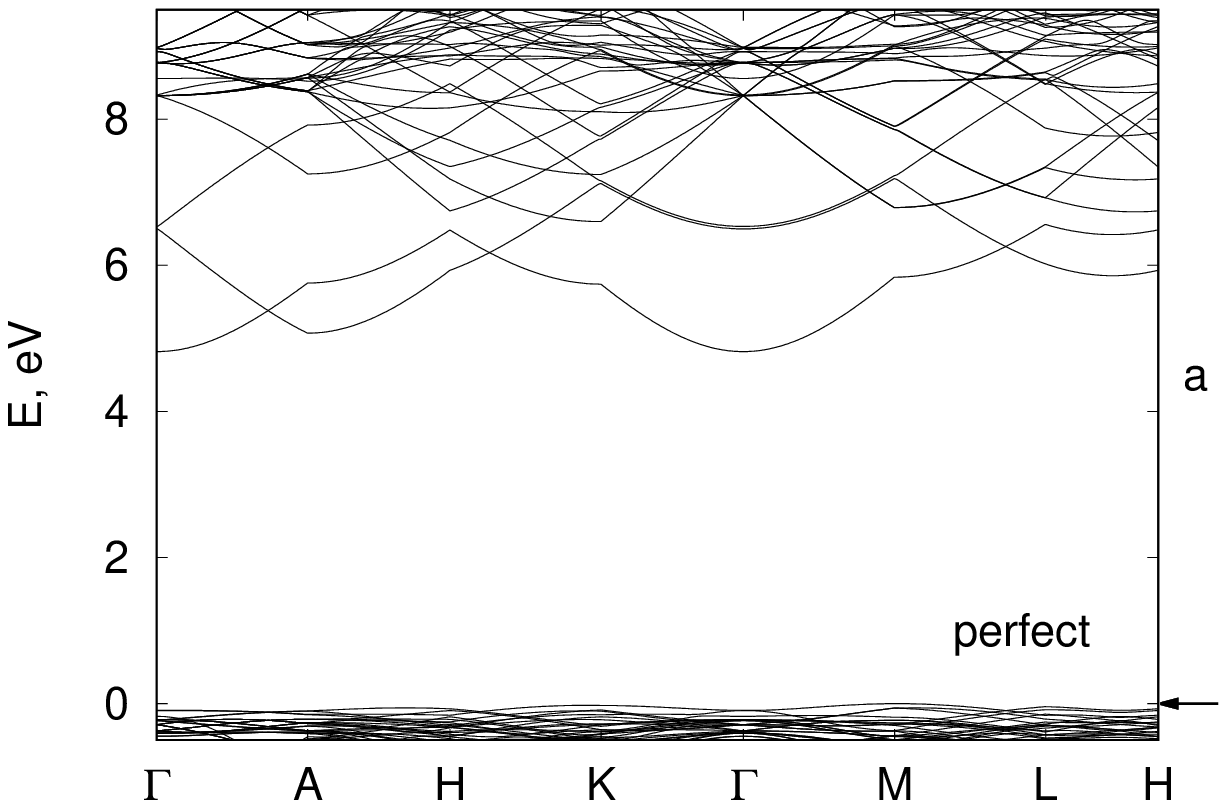}}
\end{minipage}
\begin{minipage}{0.7\linewidth}
\center{\includegraphics[width=1\linewidth]{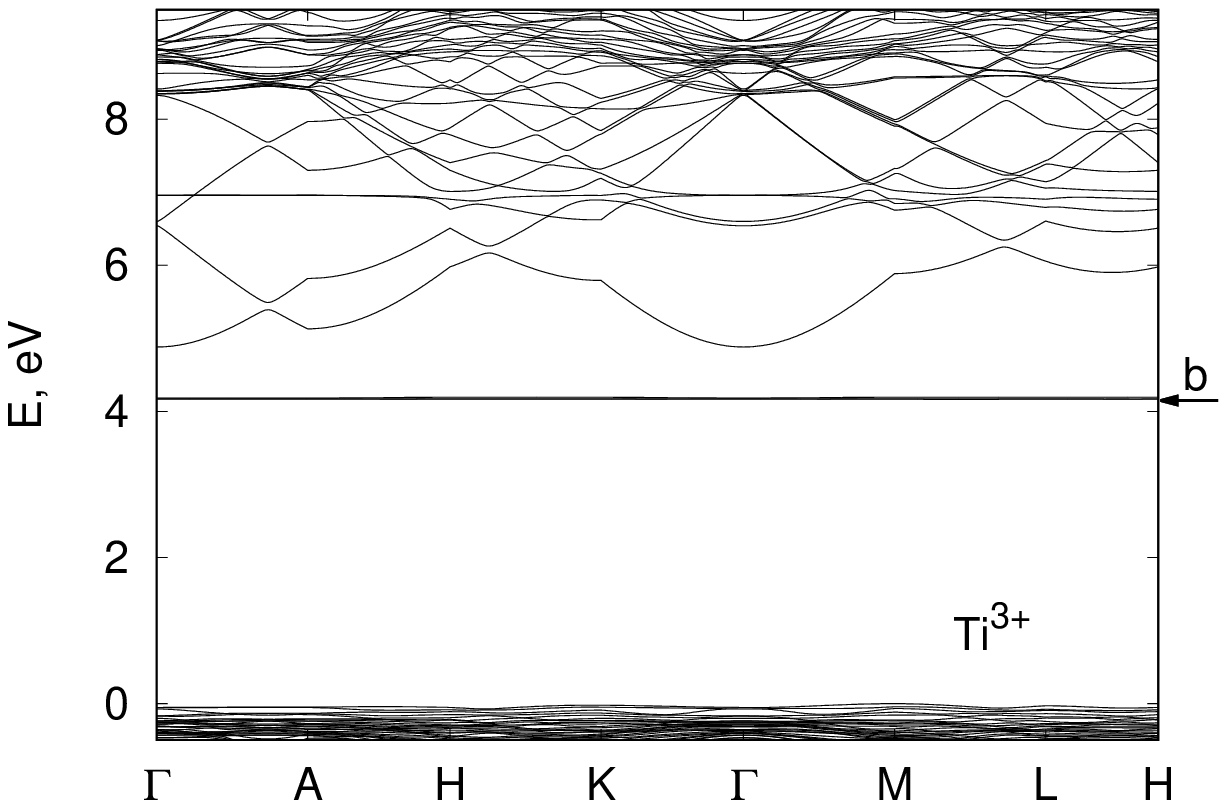}}
\end{minipage}
\begin{minipage}{0.7\linewidth}
\center{\includegraphics[width=1\linewidth]{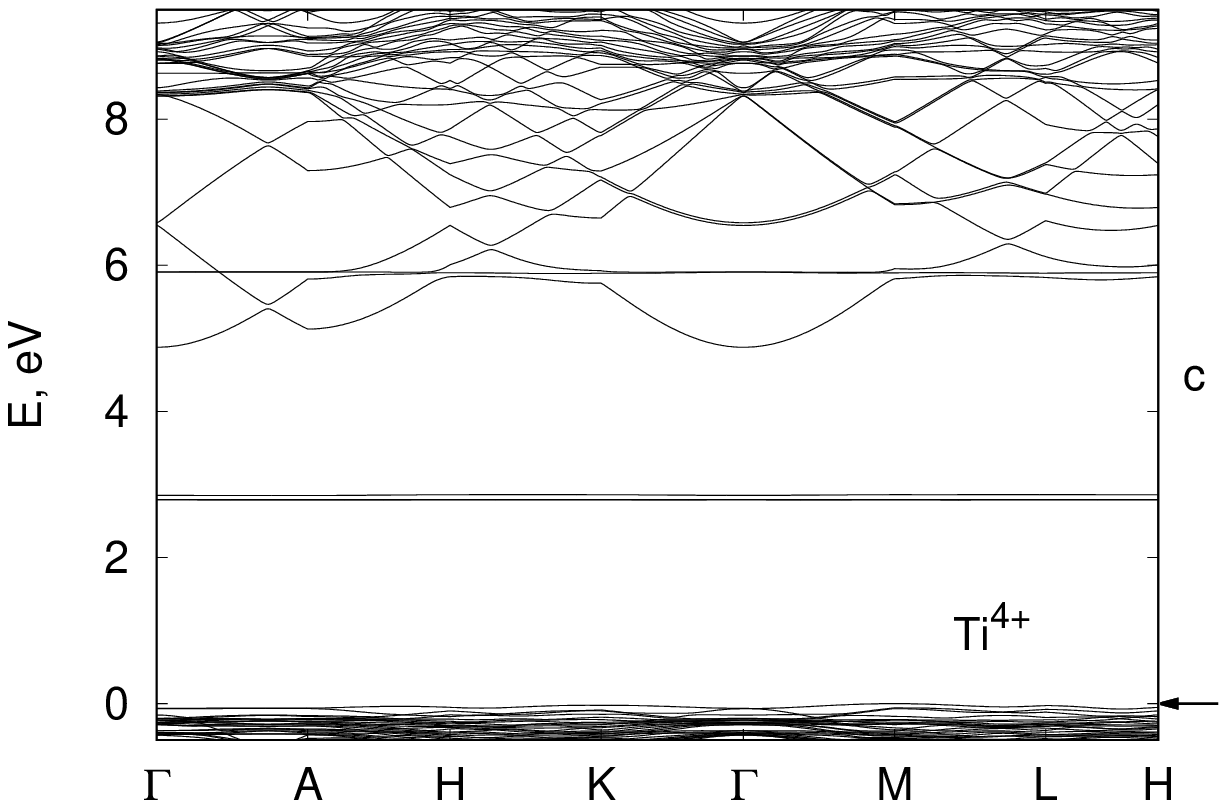}}
\end{minipage}
\begin{minipage}{0.7\linewidth}
 \center{\includegraphics[width=1\linewidth]{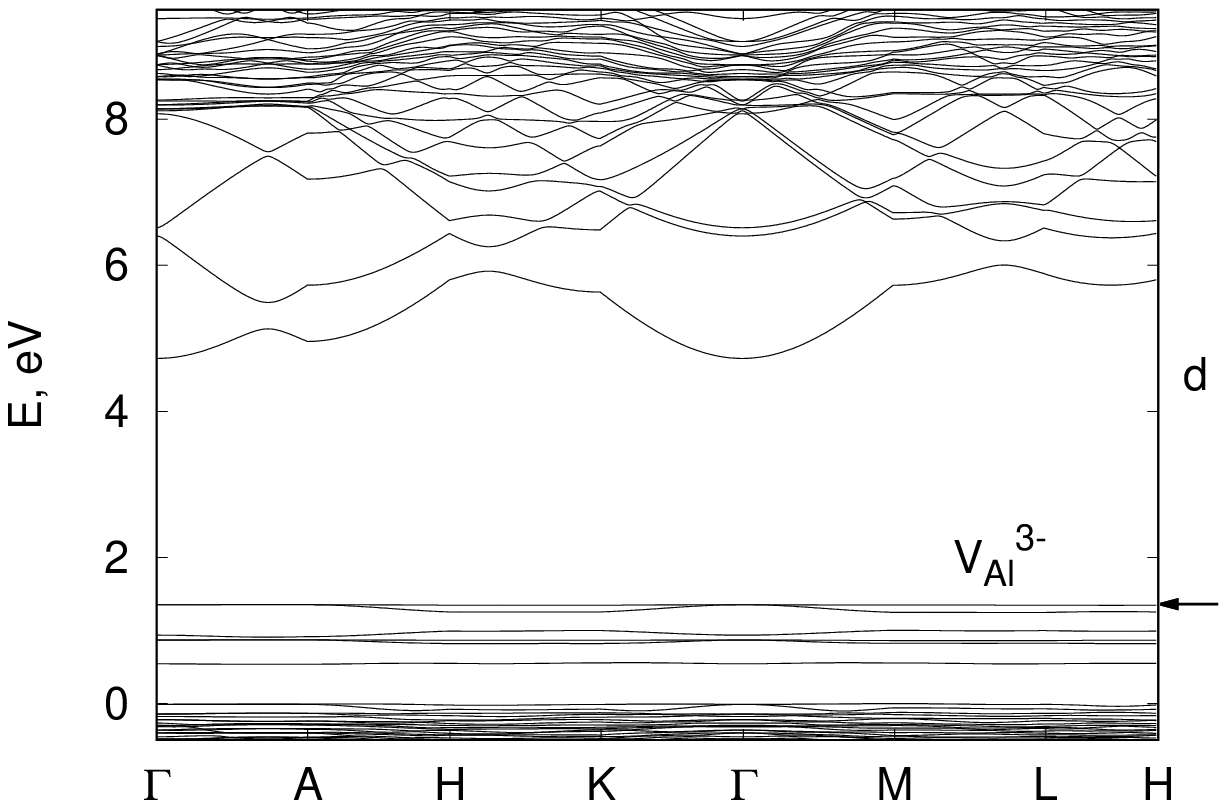}}
\end{minipage}
\caption{Band structure of the perfect $\alpha-\mathrm{Al}_2\mathrm{O}_3$ (a)  and
$\alpha-\mathrm{Al}_2\mathrm{O}_3$ with one isolated defect (b,c,d)  per supercell. The
valence band maximum is set at 0 eV, and the arrow indicates the position of the highest
occupied level.} \label{f9}
\end{figure}

In  Fig. \ref{f9}b and \ref{f9}c we present the calculated band structures of
$\alpha-\mathrm{Al}_2\mathrm{O}_3$ with one Ti$^{3+}$ and one Ti$^{4+}$ per supercell,
correspondingly. One can see that isolated Ti defects reveal itself in an appearance of
two impurity levels which correspond to $t_{2g}$ (lower) and $e_g$ (higher) states. For
Ti$^{4+}$  the lower level is much closer to the valence band maximum (VBM) than for
Ti$^{3+}$  (2.86 eV against 4.22 eV).  Similar results were obtained in \cite{14}. The
energy separation between the lower and the higher level is larger for Ti$^{4+}$ than for
Ti$^{3+}$ (3.11 eV against 2.79 eV). The additional splitting of $t_{2g}$ level is small
(0.01 eV for Ti$^{3+}$ and 0.06 eV for Ti$^{4+}$). Our calculations overestimate the
separation between $e_g$ and $t_g$ levels in comparison with experimental data \cite{28}
(2.37 eV). The positions and separations of impurity levels  shown in Fig. \ref{f7}b are
similar to ones obtained in \cite{29,30} for larger Ti concentration. Isolated Al
vacancies (Fig. \ref{f9}d) reveal itself in  the appearance of oxygen levels inside the
band gap near the VBM.

In Fig. \ref{f10} the band structure of $\alpha-\mathrm{Al}_2\mathrm{O}_3$  with  defect
pairs is presented. The a,b and c panels of Fig. \ref{f10} correspond to
$\mathrm{Ti}^{3+}-\mathrm{Ti}^{3+}$, $\mathrm{Ti}^{3+}-\mathrm{Ti}^{4+}$ and
$\mathrm{Ti}^{4+}- V_\mathrm{Al}^{3-}$ pairs, respectively. We consider the
configurations with the lowest
 formation energy.  One Ti atom or Al vacancy is
located at the central site (Fig. \ref{f2}) and another Ti atom, at the  apex site
belonging to the tetrahedron base. One can see from the comparison of Figs. \ref{f9} and
\ref{f10}  that the formation of defect pairs results in splitting of Ti impurity levels.

 In the presence of $\mathrm{Ti}^{3+}-\mathrm{Ti}^{3+}$ pairs (Fig. \ref{f10}a) the energy distance
between the lowest occupied level and the 6-th level is 2.43 eV (510 nm). It is larger
than one obtained in \cite{15}. We connect this discrepancy with the following. In \cite{15}
the band structure was calculated for the Ti$^{3+}$–Ti$^{3+}$ pair oriented along the [0001] axis while another orientation
of that pair was implied in our calculations. 
 The presence of  $\mathrm{Ti}^{3+}-\mathrm{Ti}^{4+}$ pairs (Fig. \ref{f10}b)
results in the appearance of a number of empty impurity levels, one of which is located
at around 1.37 eV (900 nm) to the occupied level. This energy distance can be related to
the experimentally observed infrared absorption peak at 800 nm.

In Fig. \ref{f11} the band structure of $\alpha-\mathrm{Al}_2\mathrm{O}_3$  with triple
and quadruple defects
 is displayed. The defects are formed by an Al vacancy at
 the central site (Fig. \ref{f2}) and  two or
 three Ti ions located at apex sites of the
tetrahedron base. It corresponds to the lowest formation energy configurations.
 Comparing Fig. \ref{f10}b and \ref{f11}a we conclude that the binding of Al
vacancies with $\mathrm{Ti}^{3+}-\mathrm{Ti}^{4+}$ pairs  results in a red shift of the
infrared absorption peak. Our calculations yield the shift $\Delta E=0.43$ eV. This shift
can result in an increase of FoM of Ti:sapphire lasers.

One can see from Figs. \ref{f9}c,  \ref{f10}c,  \ref{f11}b and \ref{f11}c  that the
lowest Ti$^{4+}$ level lifts up relative VBM under the binding of $\mathrm{Ti}^{4+}$ with
$V_\mathrm{Al}^{3-}$. The additional higher levels also appear under such a binding. The
latter may influence only insignificantly  on characteristics of Ti:sapphire lasers.

\begin{figure}
\begin{minipage}{0.7\linewidth}
\center{\includegraphics[width=1\linewidth]{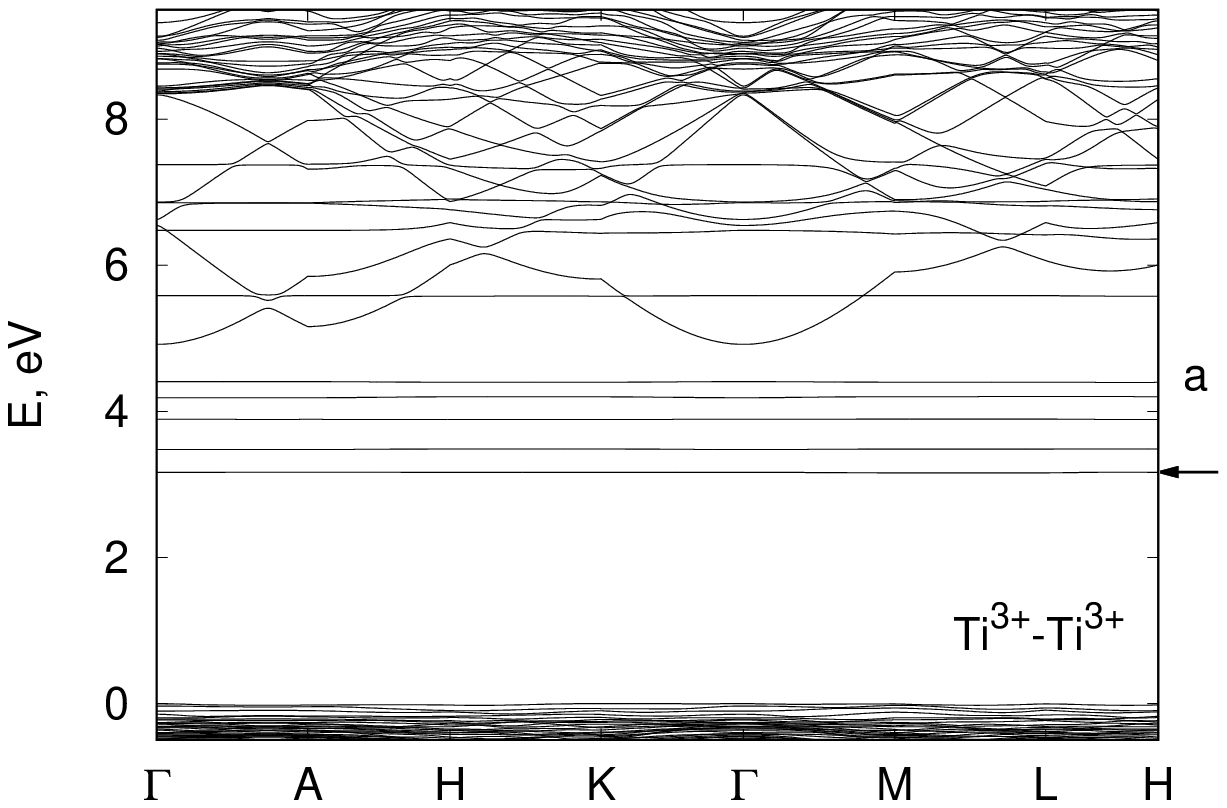}}
\end{minipage}
\begin{minipage}{0.7\linewidth}
\center{\includegraphics[width=1\linewidth]{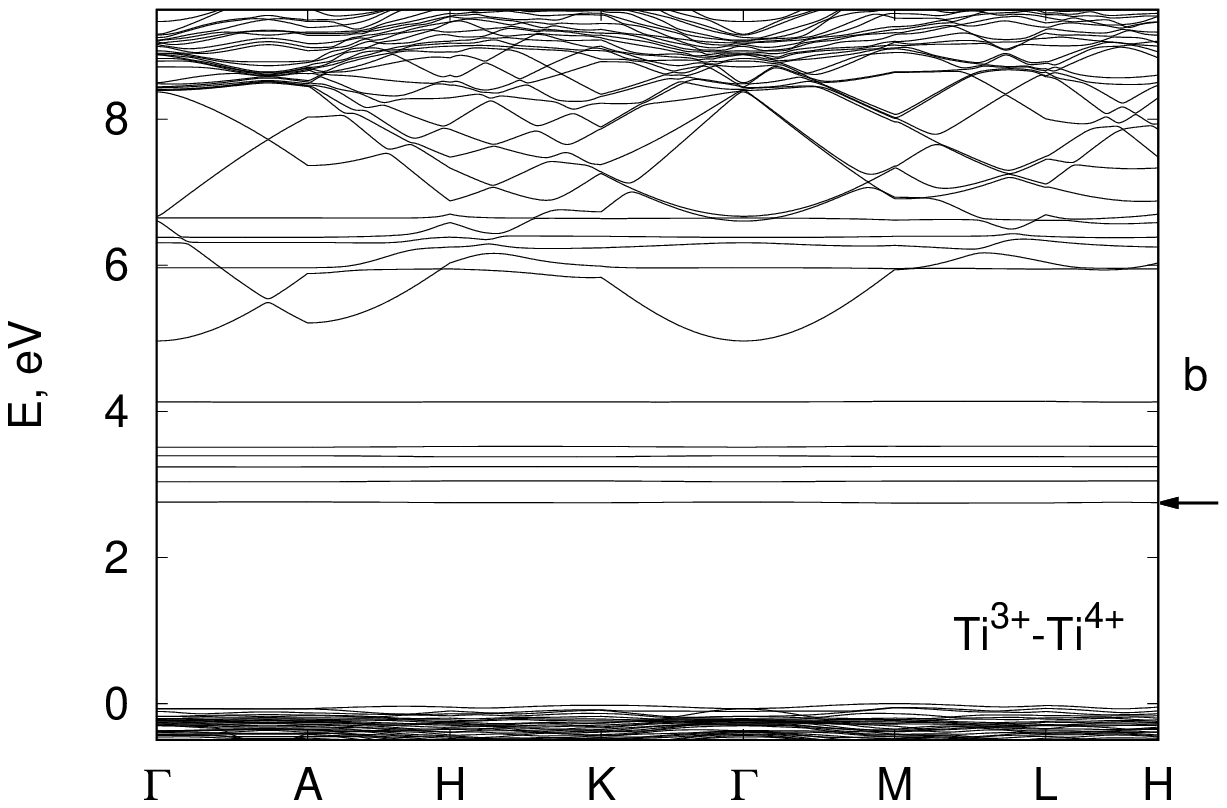}}
\end{minipage}
\begin{minipage}{0.7\linewidth}
\center{\includegraphics[width=1\linewidth]{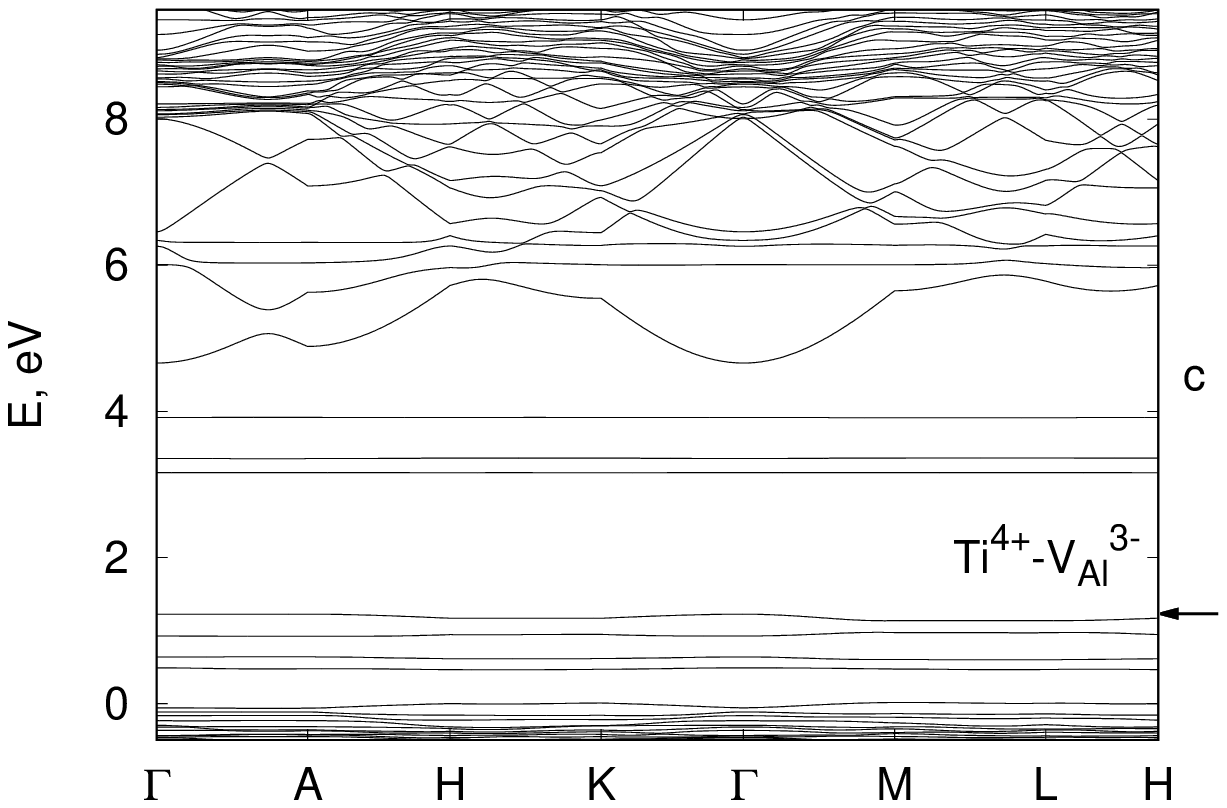}}
\end{minipage}
\caption{Band structure of $\alpha-\mathrm{Al}_2\mathrm{O}_3$ with one defect pair
 per supercell.} \label{f10}
\end{figure}

\begin{figure}
\begin{minipage}{0.7\linewidth}
\center{\includegraphics[width=1\linewidth]{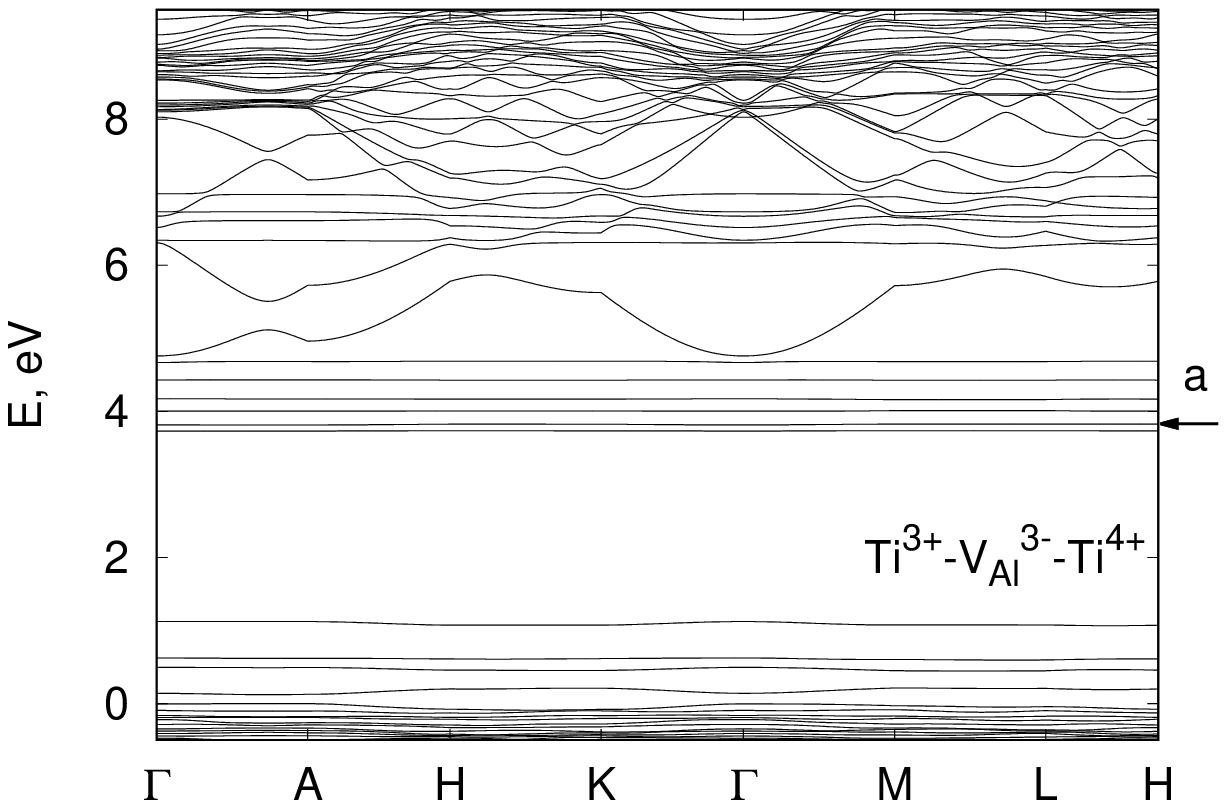}}
\end{minipage}
\begin{minipage}{0.7\linewidth}
\center{\includegraphics[width=1\linewidth]{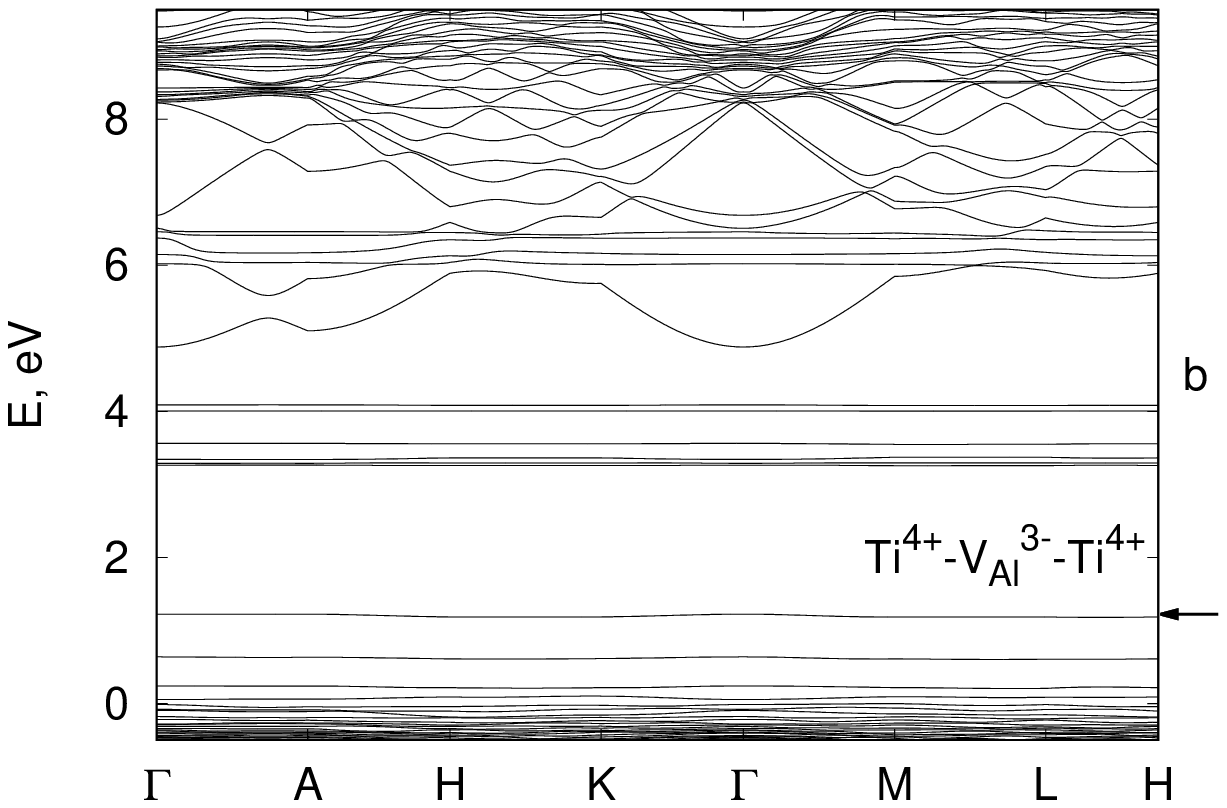}}
\end{minipage}
\begin{minipage}{0.7\linewidth}
\center{\includegraphics[width=1\linewidth]{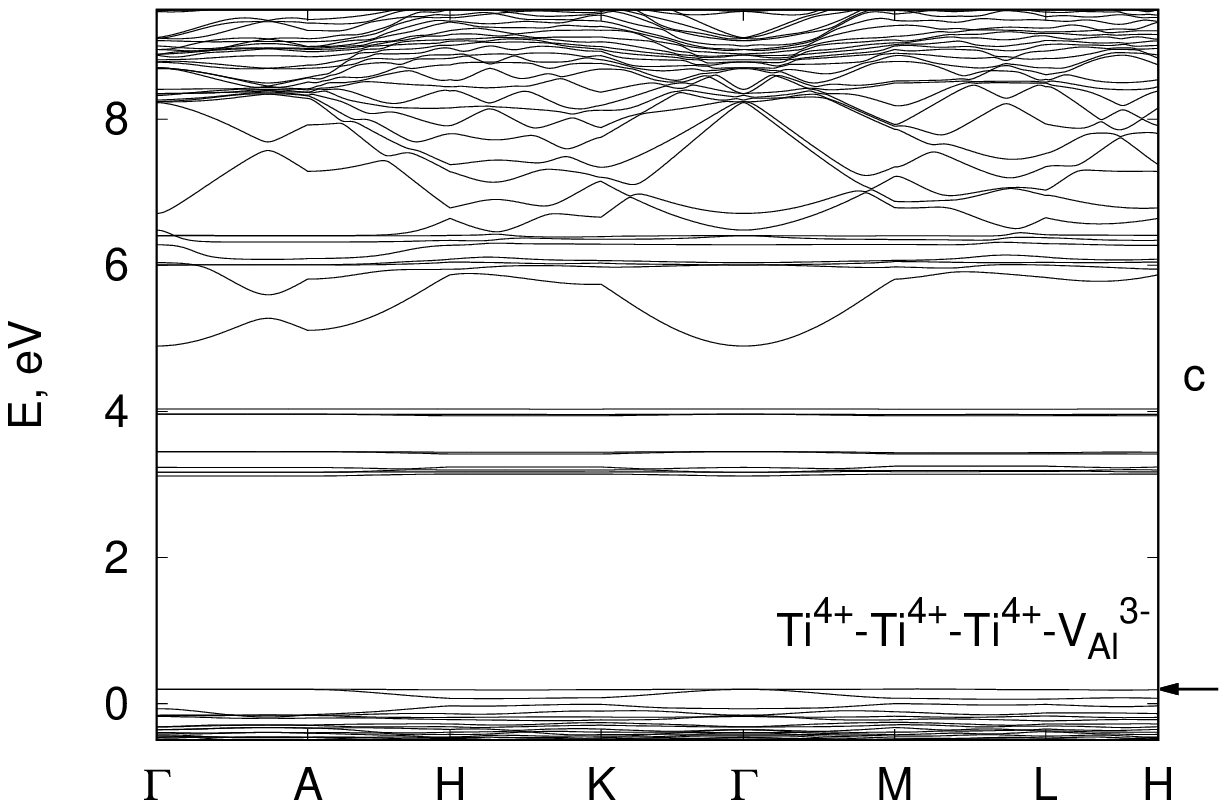}}
\end{minipage}
\caption{Band structure of $\alpha-\mathrm{Al}_2\mathrm{O}_3$ with one complex (triple or
quadruple) defect per supercell.} \label{f11}
\end{figure}

      \section{Conclusion}

In conclusion we have found by the first-principles calculations that Ti-doped
$\alpha$-Al$_2$O$_3$ contains a  large fraction of complex defects (pair, triples,
quadruples) formed by $\mathrm{Ti}^{3+}$ and $\mathrm{Ti}^{4+}$ substitutional ions and
Al vacancies $V_{\mathrm{Al}}^{3-}$. Partial   concentrations of these defects depend on
the oxygen chemical potential. A significant fraction of complex defects formed by one,
two or three $\mathrm{Ti}^{4+}$ ions and one $V_{\mathrm{Al}}^{3-}$ vacancy emerges in
the oxidized and intermediate conditions, while a large fraction of $\mathrm{Ti}^{3+}$
pairs appears in the reduced conditions. Our calculations yield a rather small fraction
of $\mathrm{Ti}^{3+}-\mathrm{Ti}^{4+}$ pairs. The concentration of such pairs reaches the
maximum in the intermediate conditions. $\mathrm{Ti}^{3+}-\mathrm{Ti}^{4+}$ pairs
demonstrate a tendency to bind in triples with $V_{\mathrm{Al}}^{3-}$ vacancies.

 Ti-deficient conditions are also analyzed. It is shown that the deficit of Ti
leads to a decrease in the complex defect fraction and to an increase in the isolated defect
fraction with the shift of the balance between $\mathrm{Ti}^{3+}$ and $\mathrm{Ti}^{4+}$
toward
 the ions with larger valence.

A universal relation between the concentrations of isolated and complex defects valid for
any total Ti concentration is obtained.

The influence of defect clustering on impurity levels inside the band gap is considered.
 It is found that the binding of
Al vacancies with $\mathrm{Ti}^{3+}-\mathrm{Ti}^{4+}$ pairs results in a red shift of the
 infrared absorption peak. At the same time the binding of Al vacancies with
$\mathrm{Ti}^{4+}$ ions may influence only insignificantly  the laser characteristics of
Ti:sapphire.

      \section*{Acknowledgments}

      This work was performed using computational facilities of the Joint
computational cluster of State Scientific Institution "Institute for Single Crystals" and
Institute for Scintillation Materials of National Academy of Sciences of Ukraine
incorporated into Ukrainian National Grid.

\end{document}